\documentclass[aps,10pt,prb,bibliography,citeautoscript,superscriptaddress,twocolumn]{revtex4-2}
\usepackage{preamble}

\begin{document}

\title{Spin-liquid-based topological qubits}
	
\author{Kai Klocke}
\thanks{These authors contributed equally to this work.}
\affiliation{Department of Physics, University of California, Berkeley, California 94720, USA}

\author{Yue Liu}
\thanks{These authors contributed equally to this work.}
\affiliation{Department of Physics and Institute for Quantum Information and Matter, California Institute of Technology, Pasadena, CA 91125, USA}

\author{G\'abor B. Hal\'asz}
%\thanks{This manuscript has been authored by UT-Battelle, LLC under Contract No. DE-AC05-00OR22725 with the U.S. Department of Energy. The United States Government retains and the publisher, by accepting the article for publication, acknowledges that the United States Government retains a non-exclusive, paid-up, irrevocable, world-wide license to publish or reproduce the published form of this manuscript, or allow others to do so, for United States Government purposes. The Department of Energy will provide public access to these results of federally sponsored research in accordance with the DOE Public Access Plan (http://energy.gov/downloads/doe-public-access-plan).}
\affiliation{Materials Science and Technology Division, Oak Ridge National Laboratory, Oak Ridge, Tennessee 37831, USA}
\affiliation{Quantum Science Center, Oak Ridge, Tennessee 37831, USA}
\author{Jason Alicea}
\affiliation{Department of Physics and Institute for Quantum Information and Matter, California Institute of Technology, Pasadena, CA 91125, USA}
\affiliation{Walter Burke Institute for Theoretical Physics, California Institute of Technology, Pasadena, CA 91125, USA}

\date{\today}

	\begin{abstract}
		
Topological quantum computation relies on control of non-Abelian anyons for inherently fault-tolerant storage and processing of quantum information.  By now, blueprints for topological qubits are well developed for electrically active topological superconductor and fractional quantum Hall platforms.  We leverage recent insights into the creation and detection of non-Abelian anyons in  electrically insulating spin systems to propose topological qubit architectures based on quantum spin liquids.  We present two types of prototype designs that enable the requisite control in a potentially scalable framework: one invokes spin liquids integrated into magnetic tunnel junction arrays, the other uses semiconductor-spin liquid hybrids.  We further identify various protocols for interrogating spin-liquid-based topological qubits, both to validate the underlying principles of topological quantum computation and to establish gates required for universal quantum computation.  These results provide long-term direction for experimental investigation of Kitaev materials and potentially other solid-state spin liquid hosts.  
 
	\end{abstract}
	\maketitle

\section{Introduction}

Topological phases of matter hosting non-Abelian anyons (or closely related non-Abelian defects) can emerge in experimental platforms including topological superconductors~\cite{Read_2000,Kitaev_2001}, fractional quantum Hall systems~\cite{Moore_1991}, and quantum spin liquids~\cite{Kitaev_2006,Jackeli_2009,Balents_2010,Savary_2016,Zhou_2017,Knolle_2019,Broholm_2020}.  Interest in these settings is fueled in large part by their potential utility for intrinsically fault-tolerant topological quantum information processing~\cite{Kitaev_2003,Nayak_2008}.  Three deeply related non-Abelian-anyon characteristics \cite{RowellWang} underlie this technological promise: First, a collection of non-Abelian anyons nucleated in a host platform encodes a degeneracy consisting of locally indistinguishable states.  Qubits encoded in that manifold correspondingly enjoy built-in resilience against dephasing errors from local environmental noise.  Second, adiabatically braiding non-Abelian anyons nontrivially rotates the system's wavefunction within the degenerate subspace---enacting a set of `rigid' quantum gates on the protected qubits.  Third, non-Abelian anyons can `fuse' into multiple quasiparticle types, with a history-dependent outcome.  Measuring their fusion products thus provides a mechanism of quantum-state readout.  

Building a functional prototype topological qubit requires definitive formation of a non-Abelian topological phase, on-demand creation and controlled manipulation of non-Abelian anyons, and detection of individual anyons and their possible fusion products.  Topological qubit blueprints integrating these requirements were first proposed for quantum Hall systems~\cite{DasSarma_2005,Bravyi_2006,Freedman_2006} and later adapted to topological superconductors~\cite{Fu_2008,Sau_2010,Alicea_2010,Lutchyn_2010,Oreg_2010,Alicea_2012,Beenakker_2013} (leading eventually to scalable designs~\cite{Karzig_2017}).  Despite the rather different nature of the two platforms, both constitute electrically active systems amenable to gating and charge transport---key experimental tools for manipulation and detection.  In the quantum Hall setting, for example, electrostatic gating can nucleate fractionally charged non-Abelian anyons that can be electrically detected via anyon interferometry~\cite{Chamon_1997,DasSarma_2005,Stern_2006,Bonderson_2006,Bonderson_2006Jul,Kim_2006,Rosenow_2007,Halperin_2011,Rosenow_2012}.  Similarly, gating can manipulate non-Abelian domain-wall defects in topological superconductors~\cite{Alicea_2011}, while electrical transport in conjunction with Coulomb charging effects enables interferometric readout~\cite{Fu_2010,Plugge_2016,Vijay_2016,Plugge_2017,Karzig_2017}.  

How does one build and interrogate a topological qubit based on a non-Abelian quantum spin liquid?  This question is nontrivial because spin liquids appear in electrically inactive spin systems.  Consequently, one can not naively import the topological qubit designs and control techniques noted above to this setting.  The question is also relevant given intense experimental investigations of a family of magnetic insulators dubbed Kitaev materials~\cite{Jackeli_2009,Winter_2017,Trebst_2017,Hermanns_2018,Janssen_2019,Takagi_2019,Motome_2019}---believed to exhibit Hamiltonians `close' to the exactly solvable Kitaev honeycomb model~\cite{Kitaev_2006}.  The Kitaev model in particular harbors a magnetic-field-induced non-Abelian spin liquid phase, for which experimental evidence has been reported in \alpRuCl{3}~\cite{Banerjee_2018,Kasahara_2018,Balz_2019,Yokoi_2021,Bruin_2022,Imamura_2024} (see, however, Refs.~\onlinecite{Bachus_2020,Yamashita_2020,Bachus_2021,Chern_2021,Czajka_2021,Lefrancois_2022,Czajka_2022}).  

Numerous recent papers have devised schemes for creating and detecting non-Abelian anyons in spin liquids~\cite{Aasen_2020,Pereira_2020,Konig_2020,Udagawa_2021,Klocke_2021,Jang_2021,Wei_2021,Klocke_2022,Liu_2022,Bauer_2023,Wei_2023,Takahashi_2023,Harada_2023,Kao_2024a,Kao_2024b,Halasz_2024} (see also the references in Sec.~\ref{Review}).  We build on these results to propose spin-liquid-based topological qubit blueprints as well as protocols designed to validate the essential underpinnings of topological quantum computation in this setting.  We develop two types of architectures that incorporate potentially scalable ingredients: (i) spin liquids integrated into magnetic tunnel junction arrays that allow local, real-time manipulation of the host platform, and (ii) semiconductor-spin liquid hybrids for which electrons bind non-Abelian anyons, enabling charged-based control of the latter.  In both cases readout proceeds via anyon interferometry.

Following a philosophy similar to Ref.~\onlinecite{Aasen2016}, we primarily focus on relatively simple geometries that allow for a series of proof-of-concept topological qubit experiments.  We specifically present device designs and protocols for qubit initialization, verification of non-Abelian anyon fusion rules, demonstration of non-Abelian braiding, and extraction of topological qubit lifetimes.  The required ingredients in turn allow one to realize both protected and unprotected quantum gates required for a universal gate set, which we also discuss in some detail drawing inspiration from earlier work in other settings~\cite{Bravyi_2002, Kitaev_2003, Bravyi_2005, DasSarma_2005, Bravyi_2006, Georgiev_2006, Freedman_2006, Georgiev_2008, Nayak_2008, Ahlbrecht2009Mar, Bonderson_2009_MBQC, Bonderson_2010, Clarke_2010, Alicea_2011, Alicea_2012, Georgiev_2016, Karzig_2017}. Our designs moreover extend to scalable multi-qubit architectures.  While many advances would be required to bring our proposals to life, we hope that the strategies developed here provide useful long-term guidance for quantum spin liquid experiments.

The remainder of this paper is organized as follows.  Section~\ref{Review} sets the stage by reviewing the phenomenology of non-Abelian Kitaev spin liquids and surveying previously introduced anyon creation and detection strategies.  Next, Sec.~\ref{AnyonManipulation} proposes anyon manipulation methods that use either magnetic tunnel junction arrays or electrical setups.  Section~\ref{QubitDesigns}  synthesizes these results to propose minimalist spin-liquid-based topological qubit designs and highlight protocols for validating non-Abelian fusion rules, non-Abelian braiding, and qubit lifetime extraction.  Section~\ref{sec:gates} adapts protocols for gates needed to achieve universal quantum computation in this setting.  Finally, we conclude in Sec.~\ref{Discussion} by highlighting various future challenges and opportunities.

\section{Review of anyon generation and detection schemes}
\label{Review}

\begin{figure}[h]
    \centering
    \includegraphics[width=\columnwidth]{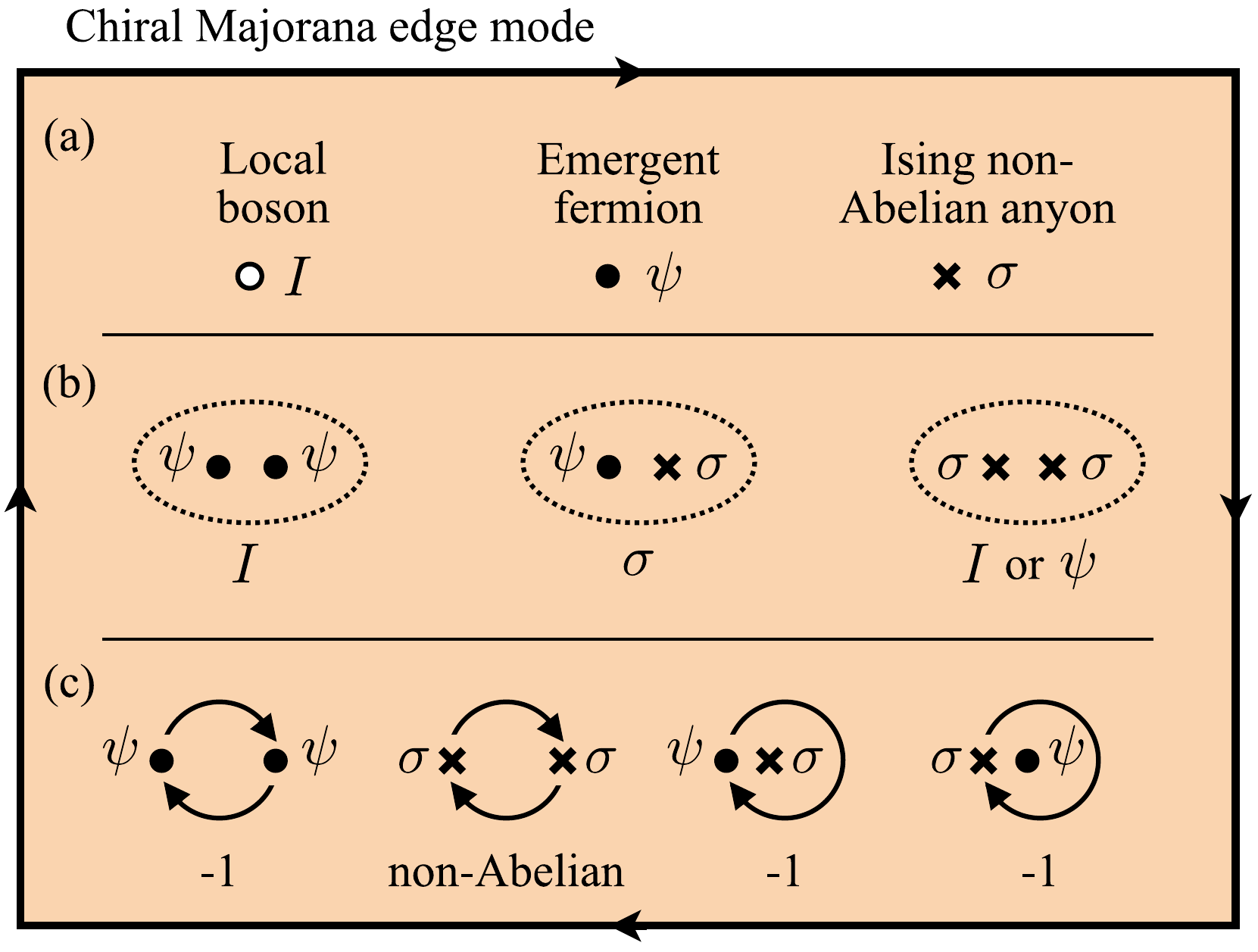}
    \caption{{\bf Topological properties of the non-Abelian Kitaev spin liquid phase.}  The boundary between the spin-liquid phase and a trivial phase hosts a chiral Majorana edge mode.  The gapped bulk supports three types of quasiparticles enumerated in (a); their nontrivial fusion rules and braiding statistics are summarized in (b) and (c), respectively.}
    \label{fig:summary}
\end{figure}

The non-Abelian Kitaev spin liquid phase hosts a gapless chiral Majorana edge mode described by a chiral Ising CFT with central charge $c=1/2$.  In the bulk, three gapped quasiparticle types can appear: topologically trivial local bosons ($I$), emergent fermions ($\psi$), and non-Abelian Ising anyons ($\sigma$) that correspond to $\mathbb{Z}_2$ flux excitations [see Fig.~\ref{fig:summary}(a)].  Non-Abelian properties of the Ising anyons descend from an emergent Majorana zero mode that binds to a $\mathbb{Z}_2$ flux in this phase.  A pair of well-separated Ising anyons (binding Majorana zero modes $\gamma_1$ and $\gamma_2$) thereby shares a single fermionic level [described by $f  \equiv (\gamma_1+i\gamma_2)/2$] that can be either empty or filled without changing the system's energy.  In turn, these zero modes encode a locally indistinguishable ground-state degeneracy that grows exponentially as the number of bulk Ising anyons increases.  

Fermions and Ising anyons obey fusion rules
\begin{align}
    \psi\times\psi=I, ~~~\psi\times\sigma=\sigma, ~~~\sigma\times\sigma=I+\psi
\end{align}
that specify how these quasiparticles behave when they coalesce [see Fig.~\ref{fig:summary}(b)].  The first fusion rule dictates that two emergent fermions combine to form a local boson. 
The second indicates that an Ising anyon that absorbs an emergent fermion remains an Ising anyon; this property reflects the fact that the absorption merely changes the occupation of the fermionic two-level system shared jointly with some distant Ising anyon.  The last fusion rule indicates that two Ising anyons can either annihilate or form an emergent fermion.  These two `fusion channels' correspond to the unoccupied and occupied states of their shared fermionic two-level system---which will no longer be degenerate when the associated Majorana zero modes are sufficiently close to hybridize appreciably.  
 
Local bosons exhibit trivial braiding statistics with each other and with all other quasiparticle types.  Emergent fermions of course exhibit fermionic statistics with each other, but also display nontrivial mutual statistics with Ising anyons: Dragging a fermion all the way around an Ising anyon or vice versa yields a statistical phase of $-1$.  Finally, Ising anyons obey non-Abelian braiding statistics.  That is, adiabatically swapping the location of Ising anyon pairs non-commutatively rotates the system's wavefunction within the ground-state manifold in a manner that depends only on topological properties of the anyon trajectories.  Figure~\ref{fig:summary}(c) summarizes the above quasiparticle braiding properties. 

We stress that local indistinguishability of the ground-state subspace, nontrivial fusion rules, and non-Abelian braiding statistics are deeply related hallmarks of non-Abelian anyons \cite{RowellWang}---which in the present context all elegantly descend from emergent Majorana zero modes bound to $\mathbb{Z}_2$ fluxes.  These hallmarks further underlie utility for intrinsically fault-tolerant quantum computation.  Harnessing this potential in Kitaev spin liquids requires (at a minimum) the ability to create, detect, and manipulate fractionalized excitations on demand in the electrically insulating host platform.  The remainder of this section reviews prior work on anyon generation and detection, some of which we will later leverage to propose potentially scalable anyon manipulation strategies and spin-liquid-based topological qubit designs.

\subsection{Detection via local probes}

Detecting individual anyons in non-Abelian Kitaev spin liquids poses a daunting task because of the limited way in which the fractionalized degrees of freedom couple to local probes.  For instance, electrons from a lead can not directly hybridize with emergent Majorana fermions born in the spin liquid (contrary to the situation for topological superconductors where Majorana fermions arise from physical electron and hole superpositions).  
A given anyonic excitation nevertheless corresponds to a localized packet of energy, which in turn generically couples to other local observables, e.g., through local magnetic correlations~\cite{Jang_2021}. Given the absence of time-reversal symmetry, the anyons most naturally couple to the local magnetization density~\cite{Mizoguchi_2020, Harada_2023}. Due to spin-orbit interaction combined with the finite charge gap of the quantum spin liquid, however, each anyon also locally induces electric polarization~\cite{Pereira_2020, Freitas_2024} and orbital currents~\cite{Banerjee_2023} (see also Refs.~\onlinecite{Buleaevskii_2008, Khomskii_2010} for the underlying mechanism), which may be experimentally observable by surface-probe techniques like scanning tunneling microscopy or atomic force microscopy. The key issue with this approach is that different anyons, or even trivial bosons, can exhibit similar signatures; any concrete route to detecting a specific anyon must therefore rely heavily on precise microscopic details and careful experimental calibration.  

Alternative proposals aim to detect the Majorana zero modes attached to individual Ising anyons. In these proposals, an atomically thin spin-liquid layer is assumed to form a tunneling barrier between two electronic systems~\cite{Carrega_2020, Konig_2020}, typically a scanning tunneling tip and a metallic substrate~\cite{Udagawa_2021}. The inelastic part of the tunneling response is then sensitive to the local spin dynamics~\cite{Feldmeier_2020}, and Majorana zero modes bound to Ising anyons are reflected in a sequence of distinctive features at the lowest voltages~\cite{Udagawa_2019, Bauer_2023, Takahashi_2023, Kao_2024a, Kao_2024b}. The feasibility of these inelastic tunneling setups is demonstrated by the recent observation of magnetic excitations in few-layer $\alpha$-RuCl$_3$ using inelastic electron tunneling spectroscopy~\cite{Yang_2023, Miao_2023}. Due to the inherently fractionalized nature of spin-liquid-based Majorana zero modes alluded to above, however, the low-energy spin dynamics never involves a single zero mode, but either a pair of hybridizing zero modes or a zero mode as well as a finite-energy mode~\cite{Kao_2024a, Kao_2024b}. Therefore, any signature of an \emph{individual} Ising anyon in this kind of detection scheme is necessarily indirect.

\subsection{Interferometric readout}\label{sec:interferometry}

Whereas local probes may reveal some signatures of individual anyons in the bulk, anyon interferometry enables measurement of the total topological charge within an extended region that may contain many anyons.
This type of more global detection is essential not only for reading out the logical qubit state encoded in bulk anyons, but also for implementing measurement-based quantum computation protocols.
Universality of the anyon braiding statistics means that interferometric readout eschews many of the difficulties associated with local probes; charge neutrality of the anyons, however, remains a challenge, rendering conventional charge-based interferometry techniques from the quantum Hall setting~\cite{Chamon_1997,DasSarma_2005,Stern_2006,Bonderson_2006,Bonderson_2006Jul,Kim_2006,Rosenow_2007,Halperin_2011,Rosenow_2012} ineffective.
Nonetheless, a number of schemes have been proposed to circumvent this issue by coupling the interfering anyons to more easily addressable degrees of freedom from which the statistical phase may be measured.
These strategies range from time-domain measurements coupling local spin probes to the chiral Majorana edge mode~\cite{Klocke_2021} to electrical conductance measurements in heterostructures that mediate a coupling between anyons and electronic degrees of freedom~\cite{Aasen_2020, Halasz_2024}.

\begin{figure}[h]
    \centering
    \includegraphics[width=\linewidth]{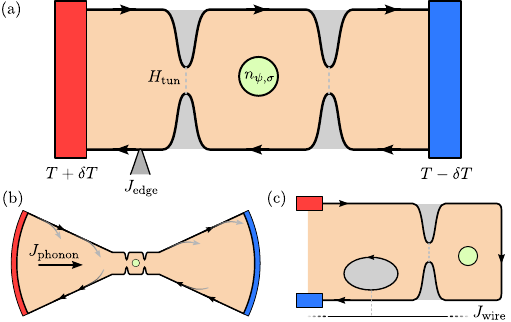}
    \caption{
    \textbf{Thermal anyon interferometry.}
    (a) Schematic thermal-transport-based readout of the topological charge in the bulk region between two quantum point contacts. A temperature gradient is induced by connections to two thermal reservoirs at temperatures $T \pm \delta T$. Tunneling at the point contacts gives rise to a shift in the edge thermal current $J_\textrm{edge}$ that depends on bulk anyon numbers $n_{\psi,\sigma}$.
    (b) In Kitaev materials with strong coupling between bulk phonons and the chiral edge mode, bulk-edge thermalization in macroscopic lobes allows for probing the bulk anyon content via phonon temperature measurements~\cite{Klocke_2022}.
    (c) Alternative proposals for thermal interferometry use \emph{local} coupling between the edge mode and, e.g., a quantum wire. Introducing a hole in the bulk near the edge may enhance the edge-wire coupling~\cite{Wei_2023}.
    }
    \label{fig:interferometry}
\end{figure}

Here we focus on a third approach---thermal anyon interferometry~\cite{Klocke_2022, Wei_2021, Wei_2023}---wherein the heat current carried by the edge mode is sensitive to the bulk topological charge.
Consider a geometry as shown in Fig.~\ref{fig:interferometry}(a) consisting of an interferometry region defined by two point contacts and flanked by two %macroscopic 
lobes.
The point contacts may either be permanently fixed by etching the edge channels or dynamically induced by, e.g., tunable magnetic tunnel junctions which bend the outer edge of the spin liquid inward to form a pinch (see Sec.~\ref{sec:DynamicalGeneration}).
Applying a temperature difference $T_{L/R} = T \pm \delta T$ across the device induces a heat current $J$, resulting in thermal conductance $\kappa = J / (2 \delta T)$.
Between the point contacts, bulk anyons encode some logical state to be read out via the total topological charge, specified by the number of Ising anyons ($n_\sigma$) and emergent fermions ($n_\psi$).
Tunneling of Ising anyons at the point contacts, with strength $t_{L/R, \sigma}$, mediates an exchange of energy between the edges and permits the chiral edge mode to acquire a statistical phase when its path encloses the interferometry region.
This braiding phase is directly reflected in thermal transport measurements as a distinct temperature-dependent correction that is sensitive to the bulk topological charge,
\begin{equation}
    \frac{\Delta \kappa }{\kappa} \sim \frac{1}{T^{7/4}}\{t^2_{L,\sigma} + t^2_{R,\sigma} + (-1)^{n_\psi}[1+(-1)^{n_\sigma}]t_{L,\sigma}t_{R,\sigma}\},
    \label{eq:thermal_conductance_correction}
\end{equation}
in addition to less relevant corrections arising from fermion tunneling~\cite{Klocke_2022}.  The piece $\propto t_{L,\sigma} t_{R,\sigma}$ represents a crucial interference term that can be understood from the braiding relations summarized in Fig.~\ref{fig:summary}(c).  For odd $n_\sigma$, the interference contribution vanishes, reflecting non-Abelian statistics of Ising anyons; that is, the paths whereby an edge Ising anyon does or does not encircle an odd number of bulk Ising anyons correspond to orthogonal states that cannot interfere.  For even $n_\sigma$, this orthogonality property is removed---hence the revival of interference, which then depends on the number of enclosed fermions due to the statistical phase $-1$ arising when an edge Ising anyon encircles a bulk fermion.

For a Kitaev material with non-negligible coupling between the edge mode and bulk phonons (e.g., \alpRuCl{3}), edge-bulk thermalization in the macroscopic lobes allows for the edge-mediated conductance to be probed via much simpler measurements of the phonon temperature [see Fig.~\ref{fig:interferometry}(b)].
Furthermore, the phonon contribution to the inter-lobe heat current can be suppressed by appropriate device geometry, making the interferometry signal $\Delta \kappa$ far more pronounced.
In this manner, thermal transport measurements may reveal the topological charge between the pinch points, thereby enabling an interferometric readout of either individual anyons or a collection of anyons with a relatively simple device design.
If the spin liquid sits upon a substrate and/or the pinch points are formed by, e.g., tunnel junctions, there are additional contributions to the thermal conductance that must be accounted for.
Nonetheless, the interferometry signal retains a unique temperature dependence.

Lastly, we note that more scalable architectures would benefit from direct measurement of the local spin temperature at the edge rather than phonon temperature in macroscopic devices.
Several schemes have been proposed for how to implement this, typically requiring more complicated heterostructures.
One such example is depicted in Fig.~\ref{fig:interferometry}(c), wherein a quantum wire is coupled to the chiral edge mode with a nearby trivial hole in the bulk making the coupling term more relevant~\cite{Wei_2023}.

\subsection{Probabilistic anyon generation via magnetic tunnel junctions}
\label{sec:DynamicalGeneration}

\begin{figure}[h]
    \centering
    \includegraphics[width=\linewidth]{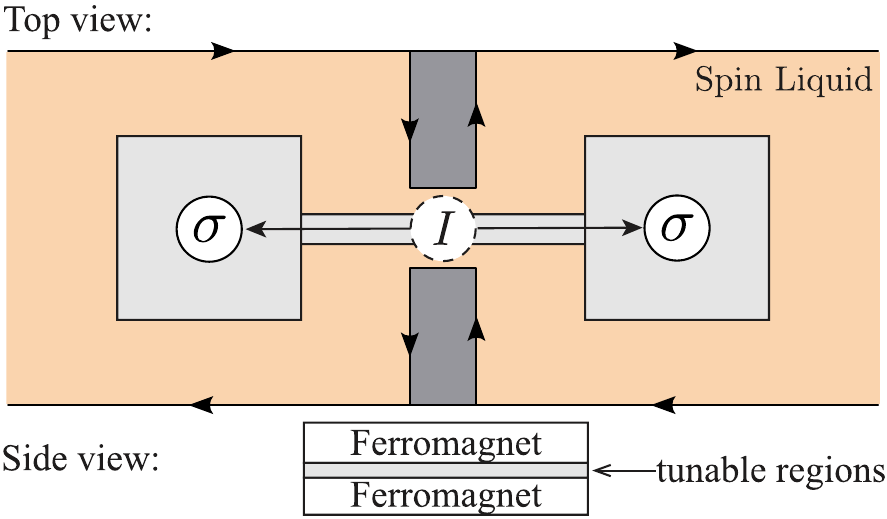}
    \caption{
    \textbf{Magnetic tunnel junction scheme for generation and readout of Ising anyons.}
    This setup consists of a non-Abelian Kitaev spin liquid (orange) punctuated by two ``holes'' of a topologically trivial phase (light gray).
    The holes are connected by a narrow ``bridge'' where the spin liquid is sandwiched between two ferromagnets, thus forming a tunnel junction.
    Dynamically switching the orientation of either ferromagnet yields control over the local Zeeman field, allowing one to tune the bridge between the topologically trivial phase and the spin-liquid phase.
    Diabatically tuning the bridge from the trivial phase to the spin-liquid phase, two Ising anyons $\sigma$ are probabilistically nucleated on the two holes.
    Further tunable regions above and below the bridge (dark gray) enable interferometric detection of either Ising anyon by bending the chiral edge mode to form a pinch point and thus allowing for, e.g., thermal anyon interferometry (see Fig.~\ref{fig:interferometry}).
    }
    \label{fig:generation}
\end{figure}

Ising anyons constitute gapped quasiparticles in the bulk of a `pure' non-Abelian spin liquid.  Starting from this clean setting, one can pursue creation of Ising anyons using two complementary strategies.  
The first applies judicious local perturbations that eliminate the energy gap for Ising anyons, such that they appear at prescribed locations in the system's minimum-energy configuration.  Atomic-scale defects, e.g., vacancies or spin impurities, have been shown to stabilize Ising anyons via this mechanism, at least within particular microscopic models~\cite{Willans_2010,Willans_2011,Dhochak_2010,Das_2016,Vojta_2016,Kao_2021,Jang_2021}. 
The second strategy seeks to trap Ising anyons as finite-energy, long-lived excitations above the ground state.  Here, the chiral Majorana edge state residing at the boundary between the spin liquid and a trivial phase provides an appealing trapping center.  The underlying CFT dictates that Ising anyons universally comprise the lowest-energy excitation for the edge state, with an energy $E_\sigma \sim \frac{1}{16} \frac{2\pi v}{L}$ that decreases as the edge circumference $L$ increases ($v$ denotes the edge velocity).  Note that away from a boundary the hierarchy of excitations can easily flip such that Ising anyons cost much larger energy than emergent fermions.  
These observations suggest the potential for dynamically generating Ising anyons by non-adiabatically altering spin-liquid edge states~\cite{Liu_2022}. One advantage of this approach is that it relies solely on universal edge physics rather than microscopic details of the host Kitaev material.

Figure~\ref{fig:generation} sketches a spin-liquid-based setup for dynamical Ising-anyon generation.  The non-Abelian spin liquid features two `holes'---both realizing a topologically trivial phase---connected by a `bridge'. Initially the bridge also resides in a trivial phase, thereby linking the two holes in a way that forms a common dumbbell-like edge state.  We assume that the temperature is sufficiently low to prevent excitations along this dumbbell edge.  Next, suppose that the bridge can be tuned from a trivial phase to the spin-liquid phase over a time scale $\tau$, after which the two holes become decoupled and thus exhibit independent edge states.  Purely adiabatic evolution would simply initialize the independent edge states into their ground state---hence no anyon generation.  Conversely, too-rapid evolution would generate unwanted excitations in the bridge region.  
Reference~\onlinecite{Liu_2022} found that intermediate $\tau$ satisfying 
\begin{equation}\label{eq:creation_timescale}
    \frac{L_b^2}{v^2}\Delta_{\rm bulk} \lesssim \tau \lesssim \frac{L_b L_h}{v^2} \Delta_{\rm bulk}
\end{equation}
generates an Ising anyon in each hole with $\mathcal{O}(1)$ probability, without producing other spurious excitations.  In Eq.~\eqref{eq:creation_timescale},
$\Delta_{\rm bulk}$ is the bulk spin-liquid gap, while $L_{b}$ and $L_h$ are the length of the bridge and the perimeter of each hole, respectively.  
One can view the process as pulling a pair of Ising anyons out of the vacuum ($I \to \sigma \sigma$) and storing one in each hole. Since the generation is inherently probabilistic, measurement of $\sigma$ through, e.g., thermal anyon interferometry is needed at the end of the protocol to collapse the wavefunction onto the desired sector.  If readout does not detect an Ising anyon, the cycle repeats until successful.  
Once created, the Ising anyons are protected by the bulk gap and will not annihilate unless the bridge reverts to the trivial phase.

Equation~\eqref{eq:creation_timescale} predicts a broad suitable $\tau$ window when $L_b \ll L_h$. The probability of successful creation, $p_{\rm suc}$, also decreases with $L_b/L_h$ (fixing all other parameters, shrinking $L_h$ raises the energy cost for $\sigma$ particles in the edge states and thus suppresses $p_{\rm suc}$).  Simulations nonetheless show that $p_{\rm suc}$ can remain $\mathcal{O}(1)$ even for $L_b \sim L_h$ \cite{Liu_2022}. The condition $L_b \ll L_h$ is therefore not strictly necessary. The width of the bridge, however, should not greatly exceed the bulk spin-liquid correlation length $\xi_{\rm bulk} \sim v_{\rm bulk}/\Delta_{\rm bulk}$ (where $v_{\rm bulk}$ is the bulk emergent fermion velocity) in order to avoid a 2D phase transition during the dynamical bridge evolution.   For a sense of scales, taking $v_{\rm bulk} \sim \qty{3d3}{\metre\per\second}$ and $\Delta_{\rm bulk} \sim \qty{5}{\kelvin}$ yields $\xi_{\rm bulk} \sim \qty{5}{\nano\metre}$; moreover,  $L_{h,b} \sim \qty{100}{\nano\metre}$ yields typical bridge timescales of $\tau \sim \qtyrange[range-units = single, range-phrase = -]{1}{10}{\nano\second}$.  

Experimentally, among the various Kitaev materials, $\alpha$-RuCl$_3$ is of particular interest because of its reported (though actively debated) non-Abelian signatures when subjected to a magnetic field of $\gtrsim \qty{7}{\tesla}$~\cite{Banerjee_2018,Kasahara_2018,Balz_2019,Yokoi_2021,Bruin_2022,Imamura_2024}.  With that platform as inspiration, we postulate that the evolution from the topologically trivial to the spin-liquid phase can be realized by changing the Zeeman field from `small' to `large' values. Local Zeeman-field modifications can potentially be achieved on the required timescales by sandwiching the region of interest between two ferromagnets to form a magnetic tunnel junction. When the two ferromagnetic moments align, the Kitaev material experiences a net exchange field that we assume locally generates a spin-liquid phase.  Conversely, in the anti-aligned configuration, we assume that cancellation of the exchange fields locally produces a trivial phase (e.g., magnetic order).  
Dynamical switching between the two configurations can be realized using the spin-transfer torque technique \cite{Ralph_2008} on nanosecond timescales \cite{Devolder_2008,Cui_2010,Grimaldi_2020}, thus meeting the requirements estimated above.

\subsection{Electrical generation and readout} \label{sec:ElecticalGeneration}

\begin{figure}[t]
    \centering
    \includegraphics[width=\linewidth]{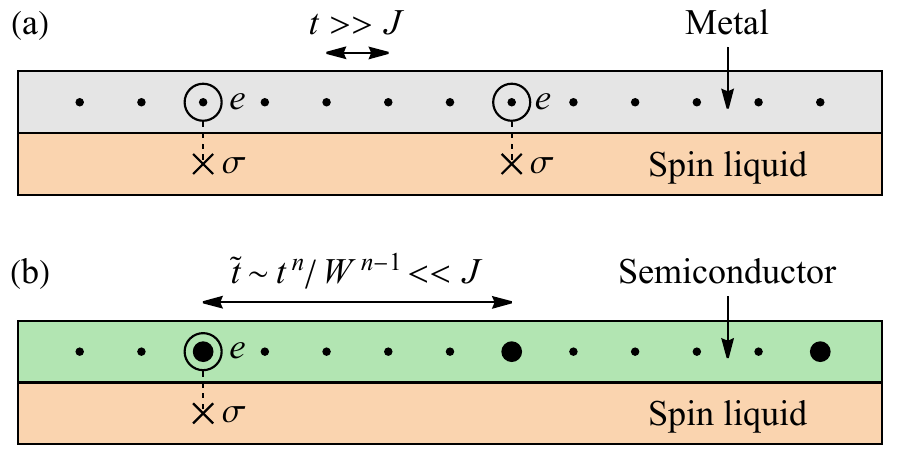}
    
    \caption{\textbf{Charge-based stabilization of Ising anyons.} (a) Naive bilayer setup consisting of a monolayer spin liquid and a monolayer metal. Each electron $e$ in the metallic layer binds an Ising anyon $\sigma$ in the spin-liquid layer due to Kondo coupling. (b) Refined setup containing a monolayer semiconductor with electron dopants (large black dots) that effectively reduce the electron hopping amplitude from $t$ to $\tilde{t}$ and hence stabilize the composite $e-\sigma$ bound states.}
    \label{fig:electric_bilayer}
\end{figure}

As an alternative route to on-demand anyon generation in a non-Abelian spin liquid, one may envision binding the anyons to electrons in a proximate metallic system, and then controlling the charged composite particles by electrical means. In particular, each electron in the proximate system can be viewed as a Kondo impurity from the perspective of the spin liquid, which---in the context of Kitaev spin liquids---is one of the localized defects proposed to stabilize non-Abelian Ising anyons~\cite{Dhochak_2010,Das_2016,Vojta_2016}. Hence, a promising platform for the electrical generation of Ising anyons is the bilayer heterostructure shown in Fig.~\ref{fig:electric_bilayer}(a) containing a single layer of a non-Abelian Kitaev spin liquid and a monolayer metal with a sufficiently strong Kondo coupling between them~\cite{Halasz_2024}.

There are two important caveats, however, that lead to the refined setup in Fig.~\ref{fig:electric_bilayer}(b). First, the large density of anyonic composites in a metallic bilayer makes generating individual anyons difficult and introduces spurious braiding between distinct anyons. Hence, one should substitute the monolayer metal with a monolayer semiconductor. Second, the stabilization of Ising anyons through a Kondo coupling has been discussed in the context of static impurity spins rather than itinerant electrons. In contrast, if the electron hopping amplitude $t$ exceeds the magnetic exchange $J$, which is the case in most real materials, the bound states of anyons and electrons are not expected to survive. To mitigate this problem, one may introduce electron dopants in the monolayer semiconductor that slow down the electron dynamics by acting like potential wells for the electrons~\cite{Halasz_2024}. Indeed, for a dopant potential depth $W>t$ and a distance $n$ between neighboring dopants (in units of the lattice spacing), the effective hopping amplitude is reduced to $\tilde{t} \sim t^n / W^{n-1}$, which can be much smaller than $J$, thus realizing a quasistatic scenario in which the anyons remain bound to the electrons as they slowly move together~\cite{Halasz_2014}. We remark that arrays of electron dopants can be created with atomistic precision in monolayer semiconductors such as transition metal dichalcogenides using the focused electron beam of a scanning transmission electron microscope~\cite{Dyck_2019}.

\begin{figure}[t]
    \centering
    \includegraphics[width=\linewidth]{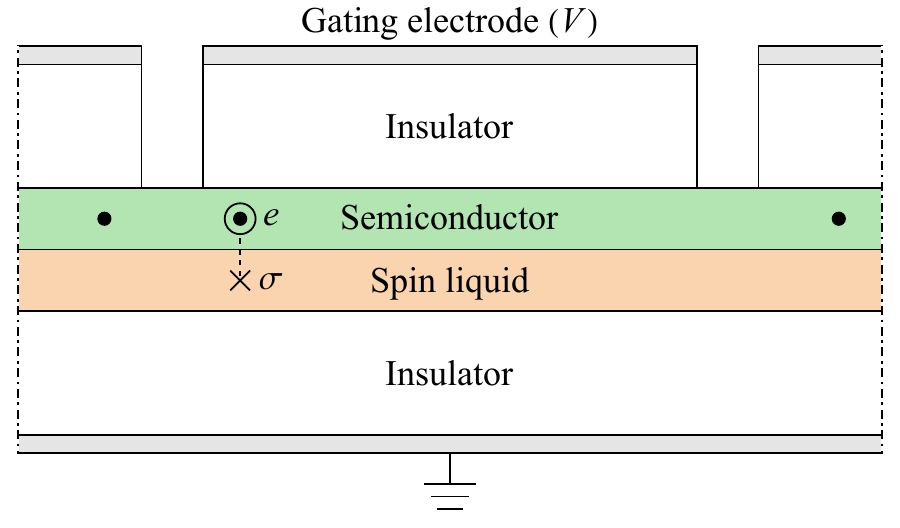}
    \caption{\textbf{Electrical generation of Ising anyons.} Numbers of electrons $e$ and, thus, Ising anyons $\sigma$ in distinct gate-defined regions are controlled by tunable gate voltages $V$.}
    \label{fig:electric_generation}
\end{figure}

The above considerations lead to the electrical anyon generation scheme in Ref.~\onlinecite{Halasz_2024} where the number of Ising anyons in each gate-defined region of the heterostructure can be controlled by tuning the appropriate gate voltage (see Fig.~\ref{fig:electric_generation}). If a given gate-defined region of the semiconducting layer---which can also be viewed as a quantum dot---is tuned into a Coulomb-blockade valley, its electron number cannot fluctuate and is fixed to a well-defined integer as long as the temperature and the effective electron hopping amplitude $\tilde{t}$ are much smaller than the charging energy. In turn, since each electron in the semiconducting layer binds an Ising anyon in the neighboring spin-liquid layer, this control over the local electron numbers enables the on-demand generation of an Ising anyon in each gate-defined region by simply tuning the appropriate electron number from zero to one.

\begin{figure}[t]
    \centering
    \includegraphics[width=\linewidth]{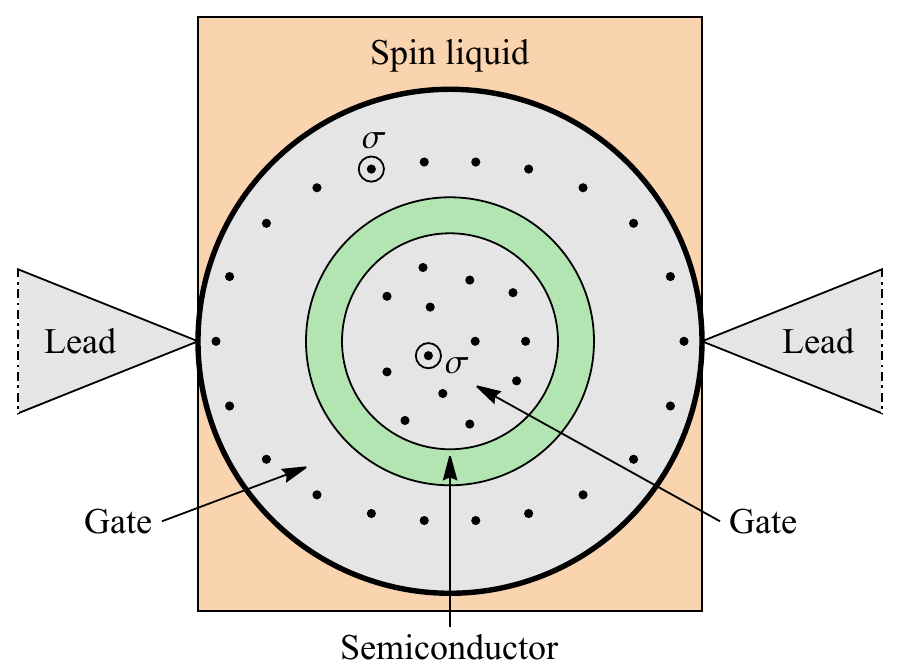}
    \caption{\textbf{Electrical readout of Ising anyons.} Top view of the heterostructure in Fig.~\ref{fig:electric_generation} with two gate-defined regions; the edge of the semiconductor is marked by a thick line. Ising anyons $\sigma$ generated in the central disk-shaped region can be interferometrically detected by measuring the electrical conductance of the surrounding ring-shaped region between two metallic leads. Note that electrons traveling between the two leads also have Ising anyons $\sigma$ bound to them.}
    \label{fig:electric_interferometry}
\end{figure}

One important advantage of the gate-controlled anyon generation approach in Ref.~\onlinecite{Halasz_2024} is being naturally interfaced with an interferometric readout scheme for detecting the Ising anyons that are stabilized in various gate-defined regions. For example, in the minimal setup shown in Fig.~\ref{fig:electric_interferometry}, an Ising anyon generated at the central disk-shaped region readily shows up in the electrical conductance of the surrounding ring-shaped region between the two metallic leads. In contrast to the disk-shaped ``generation'' region, the ring-shaped ``readout'' region is tuned to a Coulomb-blockade peak so that its electron number can easily fluctuate between zero and one. This way, the ring-shaped region has a sizable electrical conductance carried by electrons traveling from one lead to the other one. Since each electron binds an Ising anyon inside the ring-shaped region and can travel either below or above the disk-shaped region, the electrical conductance contains an interference term between the two paths that is sensitive to Ising anyons stabilized at the disk-shaped region via braiding rules of the non-Abelian Kitaev spin liquid, i.e., an emergent Aharonov-Bohm effect.

\section{Manipulation of non-Abelian anyons}
\label{AnyonManipulation}

\subsection{Magnetic tunnel junction approach}

Magnetic tunnel junctions provide a potentially scalable mechanism not only for dynamically generating Ising anyons as outlined in Sec.~\ref{sec:DynamicalGeneration}, but also for manipulating the anyons once formed.  Manipulation requires an Ising anyon to reside within an array of independently tunable magnetic tunnel junctions that can toggle the adjacent magnetic insulator regions between trivial and spin liquid phases.  Importantly, although the anyon generation protocol requires tuning between these phases on intermediate time scales---recall Sec.~\ref{sec:DynamicalGeneration}---subsequent manipulation steps proceed adiabatically to avoid generating unwanted additional excitations.  

\begin{figure}[t]
    \centering
    \includegraphics[width=\linewidth]{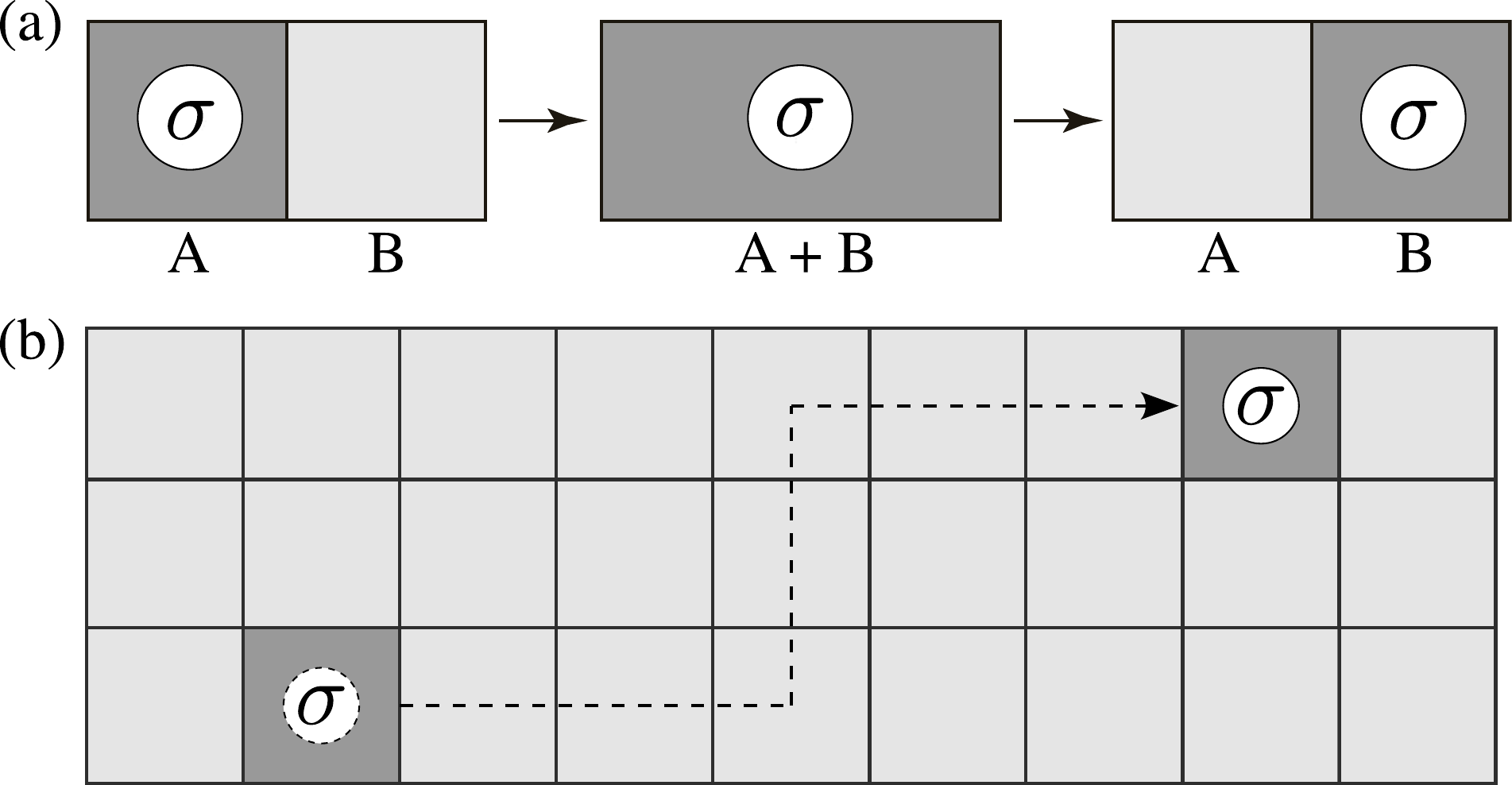}
    \caption{\textbf{Manipulation of Ising anyons in the magnetic tunnel junction scheme.} (a) Elementary `inchworm move' for transporting an Ising anyon $\sigma$ between two distinct magnetic tunnel junctions (gray squares). In each tunnel junction, the magnetic insulator is tuned to the trivial (dark gray) or the spin-liquid (light gray) phase. The individual images depict snapshots with adiabatic evolution in between. (b) In a scalable array of magnetic tunnel junctions, an Ising anyon can be transported arbitrarily far by combining elementary inchworm moves.
    }
    \label{fig:manipulation}
\end{figure}

As a minimalist example, consider first two adjacent tunable regions $A$ and $B$, initially with $A$ in the trivial phase and $B$ in the spin-liquid phase.  Suppose further that an Ising anyon is trapped in the chiral Majorana edge state surrounding $A$.  We can transport the anyon to region $B$ using the `inchworm move' sketched in Fig.~\ref{fig:manipulation}(a): First, region $B$ is adiabatically tuned from the spin-liquid to the trivial phase.  At the end of this step the spin liquid edge state encircles $A+B$, and hence the Ising anyon is delocalized across the perimeter of both regions.  Second, adiabatically tuning $A$ from trivial to spin liquid completes the migration of the Ising anyon to region $B$.  Since the edge-state excitation energy scales like $\sim v/L$ where $L$ is the characteristic size of each region, adiabaticity requires that each step occurs over a time scale that exceeds $\mathcal{O}\left(\tfrac{L^2}{v^2} \Delta_{\rm bulk}\right)$, similar to the left-hand side of Eq.~\eqref{eq:creation_timescale}.  This elementary move adapts straightforwardly to allow controlled manipulation of Ising anyons in more elaborate magnetic tunnel junction arrays; see, e.g., the manipulation sketched in Fig.~\ref{fig:manipulation}(b).

\subsection{Electrical approach}

\begin{figure}[t]
    \centering
    \includegraphics[width=\linewidth]{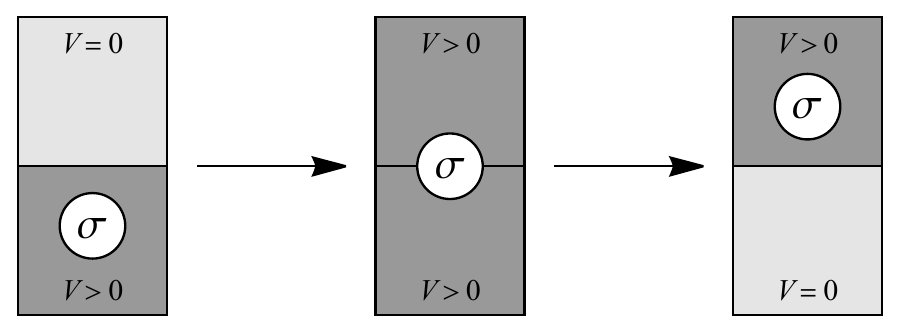}
    \caption{\textbf{Electrical manipulation of Ising anyons.} Elementary 'inchworm move' for transporting an Ising anyon $\sigma$ between two distinct gated islands (gray squares). Each gated island is a gate-defined region of the heterostructure setup in Figs.~\ref{fig:electric_generation} and \ref{fig:electric_interferometry} with a gate voltage $V$ that is tunable to $V>0$ (dark gray color) or $V=0$ (light gray color). The individual images depict snapshots with adiabatic evolution in between.}
    \label{fig:electric_manipulation}
\end{figure}

Just like its magnetic counterpart above, the electrical anyon generation scheme in Sec.~\ref{sec:ElecticalGeneration} can be straightforwardly generalized to also allow for the manipulation of the anyons generated. Once again, we envision an array of independently tunable regions [cf.~Fig.~\ref{fig:manipulation}(b)] and an elementary `inchworm move' to transport an Ising anyon between two neighboring regions $A$ and $B$. In this case, however, the regions are defined by gating electrodes and controlled by tunable gate voltages $V$ (see Fig.~\ref{fig:electric_generation}). Being attached to an electron, the Ising anyon can be localized at any given region $A$ that has a positive voltage $V>0$ with respect to the surrounding regions. Hence, in order to move the Ising anyon from region $A$ to region $B$, we first adiabatically increase the voltage at region $B$ from $0$ to $V$ (where $0$ is a default reference value), then decrease the voltage at region $A$ from $V$ to $0$. This inchworm move shown in Fig.~\ref{fig:electric_manipulation} is entirely analogous to its counterpart in the magnetic approach [see Fig.~\ref{fig:manipulation}(a)].

Importantly, it does not matter what the precise values of the voltages are during any step of the process. Indeed, while the equilibrium electron number is $N=CV/e$ for each gated island (i.e., gate-defined region) of voltage $V$ and capacitance $C$, the electron number cannot reach equilibrium as the surrounding islands of zero voltage---which have equilibrium electron numbers $N=0$---act like an energy barrier for the electrons. More precisely, if the effective electron hopping amplitude $\tilde{t}$ is much smaller than both the voltage energy $eV$ and the charging energy $e^2/C$, the tunneling of electrons in and out of the gated island is exponentially suppressed, and the lowest-energy state with exactly one electron---hence, one Ising anyon---can survive as a metastable state up to arbitrarily long time scales, regardless of the precise voltage $V$.

\begin{figure}[t]
    \centering
    \includegraphics[width=\linewidth]{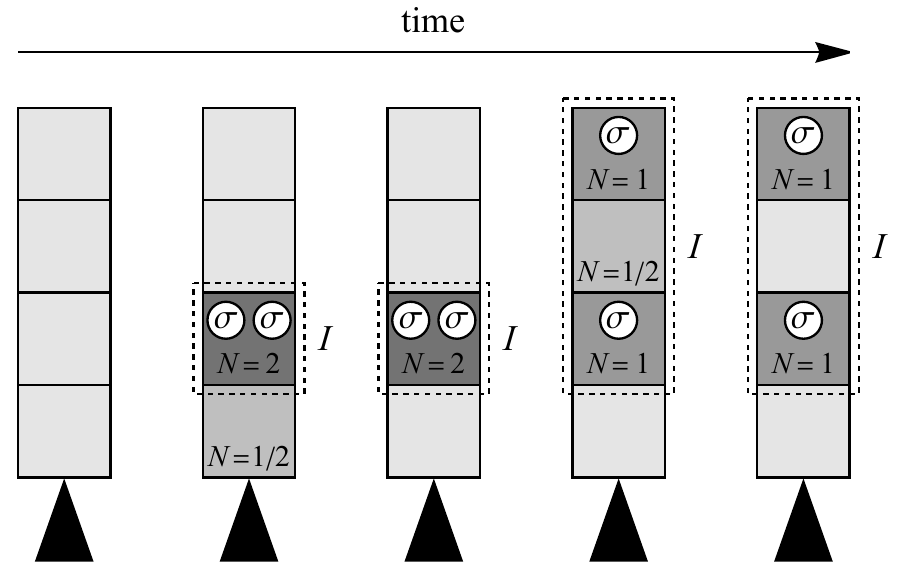}
    \caption{\textbf{Deterministic creation of an Ising-anyon pair in the electrical approach.} The individual images depict snapshots with adiabatic evolution in between. For each snapshot, the gray squares denote gated islands, while the black triangle marks a metallic lead; equilibrium electron numbers $N$ are specified for gated islands with $N>0$, and larger values of $N$ are highlighted by darker shades of gray. The details of the process are explained in the text.}
   \label{fig:electric_initialization}
\end{figure}

For the same reason, however, one must initially generate each pair of Ising anyons carefully to ensure that the system can actually reach its equilibrium state in which (i) two electrons are at distinct gated islands, (ii) each electron binds exactly one Ising anyon, and (iii) the two Ising anyons are in the trivial fusion channel $I$. Figure~\ref{fig:electric_initialization} illustrates a potential procedure for electrically pulling a pair of Ising anyons out of the vacuum, similar in spirit to the probabilistic dynamical anyon generation scheme reviewed in Sec.~\ref{sec:DynamicalGeneration} (though here the process is deterministic!). In the first step, two Ising anyons are generated on the same gated island---the `target island'---that is connected to a metallic lead by another gated island---the `bridge island'. The target island is adjusted to equilibrium electron number $N = 2$, with each of the two electrons stabilizing one Ising anyon. At the same time, the bridge island is set to $N = 1/2$, so that it can facilitate the tunneling of electrons from the lead to the target island; afterwards, the bridge island is adiabatically tuned to $N = 0$, so that it can provide an obstacle to further electron tunneling. In the second step, the two Ising anyons are separated to two different target islands that are connected by another bridge island. The target islands are both adjusted to $N = 1$, each supporting one electron (i.e., one Ising anyon) in equilibrium, while the bridge island in between facilitates the equilibration; it is tuned to $N = 1/2$ and then to $N = 0$, precisely as in the first step above. We note that the metallic lead must be far away from the spin-liquid edge to ensure that the two Ising anyons $\sigma$ are indeed pulled out of the vacuum and thus initialized in the trivial fusion channel $I$.

\section{Spin-liquid qubit designs and protocols}
\label{QubitDesigns}

\subsection{Minimal qubit architectures}

\begin{figure}[t]
    \centering 
    \includegraphics[width=\linewidth]{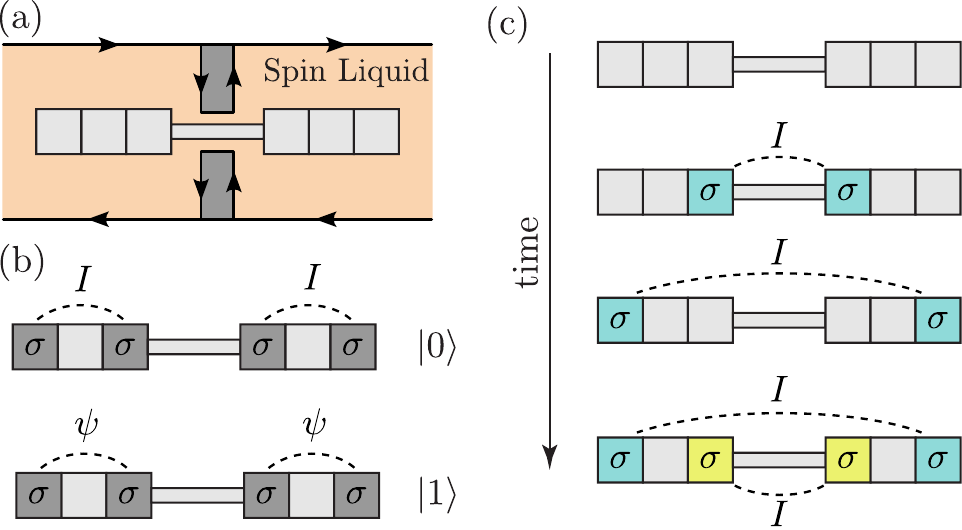}
    \caption{\textbf{Minimal qubit design and fusion experiment in the magnetic tunnel junction setup.} (a) Minimal set of tunable regions defining a single topological qubit. (b) Encoding the qubit into four Ising anyons $\sigma$ localized at regions tuned to the trivial phase (dark gray color). The two degenerate states $\ket{0}$ and $\ket{1}$ differ in fusion channels, $I$ or $\psi$, between pairs of Ising anyons $\sigma$ connected by dashed lines. (c) Qubit initialization protocol for setting up the qubit in the superposition state $(\ket{0} + \ket{1}) / \sqrt{2}$. Repeatedly implementing this initialization protocol, one can statistically demonstrate the nontrivial fusion rule $\sigma \times \sigma = I + \psi$, as described in the text.
    }
    \label{fig:fusion}
\end{figure}

\begin{figure*}[t]
    \centering
    \includegraphics[width=\linewidth]{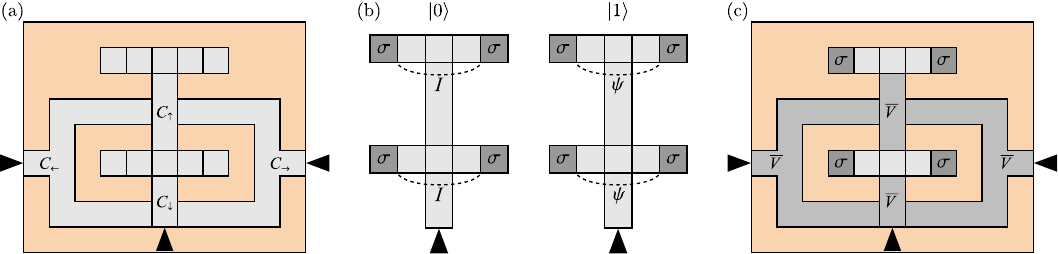}
    \caption{\textbf{Minimal qubit design in the electrical approach.} (a) Minimal set of gated islands (gray regions) and metallic leads (black triangles) required for a single topological qubit. Each gated island is a gate-defined region of the heterostructure setup in Figs.~\ref{fig:electric_generation} and \ref{fig:electric_interferometry} with its own gate voltage $V$ and capacitance $C$. Some gated islands have their capacitances specified. (b) Encoding of a single topological qubit into four Ising anyons $\sigma$ localized at gated islands with positive gate voltages (dark gray regions). The two degenerate states $\ket{0}$ and $\ket{1}$ differ in fusion channels, $I$ or $\psi$, between pairs of Ising anyons $\sigma$ connected by dashed lines. (c) Interferometric readout. Tuning the four islands around the bottom two Ising anyons to the same voltage, $\bar{V} \equiv e / [2 (C_{\uparrow} + C_{\downarrow} + C_{\rightarrow} + C_{\leftarrow})]$, corresponding to a Coulomb-blockade peak, and measuring the electrical conductance between the left and right leads, the fusion channel of the bottom two Ising anyons is interferometrically detected.}
    \label{fig:electric_qubit}
\end{figure*}

In this section, we propose a series of spin-liquid qubit architectures---focusing on minimalist designs---that synthesize the creation, manipulation, and readout techniques discussed in previous sections. The simplest architectures for encoding a single topological qubit using the two respective approaches above are shown in Figs.~\ref{fig:fusion}(a) and \ref{fig:electric_qubit}(a), with the corresponding definitions of the degenerate qubit levels $\ket{0}$ and $\ket{1}$ specified in Figs.~\ref{fig:fusion}(b) and \ref{fig:electric_qubit}(b). In general, the encoding of a single topological qubit requires not two but four non-Abelian Ising anyons $\sigma$ as accommodated in both designs. Indeed, if only two Ising anyons are nucleated from the vacuum, they necessarily fuse to the identity channel $I$, thus leaving no degeneracy to be exploited. In contrast, four Ising anyons support two degenerate ground states that can be distinguished by the fusion channel $I$ or $\psi$ of two arbitrarily chosen Ising anyons. This fusion channel of the first two anyons is identical to the fusion channel of the remaining two anyons [see Figs.~\ref{fig:fusion}(b) and \ref{fig:electric_qubit}(b)], so that the four anyons altogether still fuse to $I$.

For the magnetic tunnel junction approach, the minimal qubit design in Fig.~\ref{fig:fusion}(a) features a single tunable bridge connecting a set of three tunable holes on the left with another set of three on the right; constrictions on the top and bottom (also configurable via magnetic tunnel junctions) enable interferometric readout.  Qubit initialization proceeds as sketched in Fig.~\ref{fig:fusion}(c): (i) first, one generates a single Ising anyon pair (shown in blue) following the probabilistic, repeat-until-success scheme summarized earlier, (ii) inchworm moves then shuttle the anyons to the outer holes, and (iii) a second anyon pair (shown in yellow) is then created opposite the bridge.  Finally, interferometric readout---which 
is needed to confirm successful anyon creation in step (iii)---initializes the qubit into a well-defined state $\ket{0}$ or $\ket{1}$. 
More precisely, given the encoding specified in Fig.~\ref{fig:fusion}(b), detecting the topological charge on, say, the right three holes will return a measurement of either $I$ or $\psi$ when the creation protocol is successful---in turn collapsing the wavefunction onto either $\ket{0}$ or $\ket{1}$ depending on the outcome.

For the electrical approach, the minimal qubit architecture in Fig.~\ref{fig:electric_qubit}(a)  contains a set of gated islands, each with its own capacitance and tunable voltage, as well as three metallic leads. As shown in Fig.~\ref{fig:electric_qubit}(b), two pairs of Ising anyons can be localized at the top and bottom rows of square-shaped islands, with the two qubit states $\ket{0}$ and $\ket{1}$ distinguished by the identical fusion channels, $I$ or $\psi$, of the individual pairs. The interferometric readout of this fusion channel is depicted in Fig.~\ref{fig:electric_qubit}(c); by tuning the four islands that form a ring around the bottom row to the same voltage, $\bar{V} = e / [2(C_{\uparrow} + C_{\downarrow} + C_{\rightarrow} + C_{\leftarrow})]$, corresponding to equilibrium electron number $N = 1/2$ (i.e., a Coulomb-blockade peak), there is a measurable electrical conductance between the left and right leads, which depends on the fusion channel of the two Ising anyons inside (see Sec.~\ref{sec:ElecticalGeneration}). We note that, in contrast to the bottom lead, which is used for creating pairs of Ising anyons (see Fig.~\ref{fig:electric_initialization}), the left and right leads that enable the interferometric readout must be directly above the edge of the spin liquid. Also, the four localized Ising anyons must be well separated from the ring islands to avoid qubit decoherence from spurious interactions with the Ising anyon traveling between the leads.

To initialize the qubit in Fig.~\ref{fig:electric_qubit}, one starts by creating a single pair of Ising anyons (see Fig.~\ref{fig:electric_initialization}). Once the first pair of Ising anyons is created, qubit initialization can proceed in two different ways [see Figs.~\ref{fig:electric_fusion}(a) and \ref{fig:electric_fusion}(b)], each creating another pair of Ising anyons with appropriate inchworm moves beforehand and afterwards. At the end of the first protocol shown in Fig.~\ref{fig:electric_fusion}(a), each pair of Ising anyons is shared between the top and bottom rows, and initializing the qubit into a well-defined state $\ket{0}$ or $\ket{1}$ thus requires readout, as in the magnetic tunnel junction approach above. In contrast, for the second protocol depicted in Fig.~\ref{fig:electric_fusion}(b), the two pairs of Ising anyons end up in the two respective rows, and the qubit is deterministically initialized in the state $\ket{0}$ without any readout.

\subsection{Fusion protocols}

As noted earlier, nontrivial fusion rules and non-Abelian statistics comprise intimately linked properties of non-Abelian anyons \cite{RowellWang}.  Probing fusion rules nonetheless entails a significant practical advantage---in particular requiring simpler geometries and manipulations compared to braiding.  In fact, repeatedly implementing the initialization protocols described above for the minimal qubit architectures from Figs.~\ref{fig:fusion}(a) and \ref{fig:electric_qubit}(a) essentially realizes an experimental demonstration of the nontrivial Ising-anyon fusion rule $\sigma \times \sigma = I + \psi$!

\begin{figure*}[t]
    \centering
    \includegraphics[width=\linewidth]{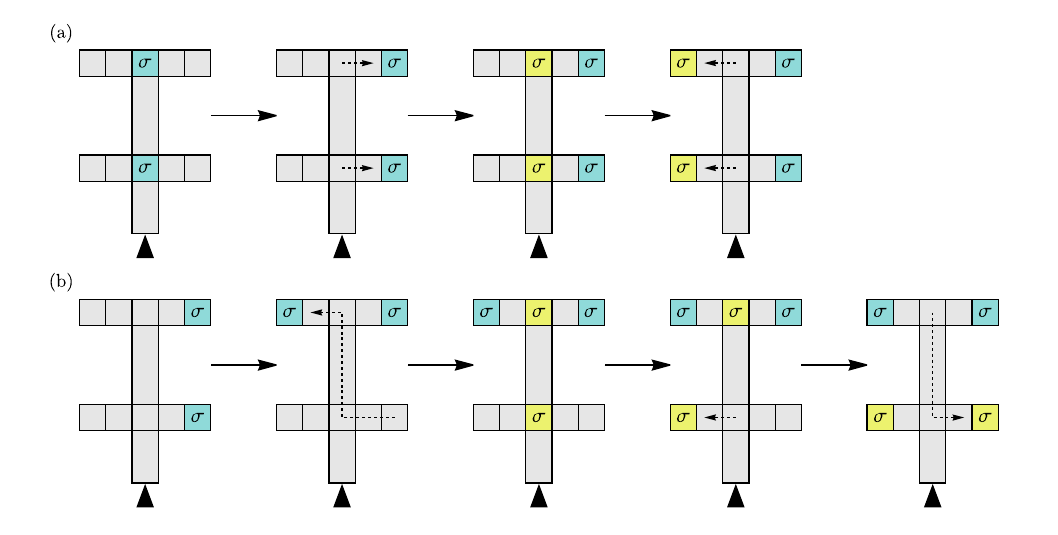}
    \caption{\textbf{Qubit initialization and fusion experiment in the electrical approach.} Two different protocols for initializing the qubit by creating two pairs of Ising anyons $\sigma$---with each of the blue and yellow pairs fusing to the identity channel $I$---and positioning them via appropriate inchworm moves (dashed arrows). Depending on the protocol, the qubit is initialized in either a superposition state $(\ket{0} + \ket{1}) / \sqrt{2}$ (a) or the basis state $\ket{0}$ (b). Repeatedly implementing these initialization protocols, one can statistically demonstrate the nontrivial fusion rule $\sigma \times \sigma = I + \psi$, as described in the text.}
    \label{fig:electric_fusion}
\end{figure*}

For concreteness, let us first focus on the magnetic tunnel junction design from Fig.~\ref{fig:fusion}(a). Suppose that the first two steps of the initialization protocol in Fig.~\ref{fig:fusion}(c) have been successfully carried out, leading to the configuration in the third row of Fig.~\ref{fig:fusion}(c) featuring an Ising anyon at each of the outermost holes.  Those anyons were effectively pulled out of the vacuum and hence reside in the identity fusion channel $I$.  Next suppose that one attempts to create a second pair of Ising anyons to arrive at the configuration in the fourth row of Fig.~\ref{fig:fusion}(c), and let $p_{\rm suc}$ denote the success probability.  When successful, the second pair generated also resides in the identity fusion channel.  Notice that the resulting initialized qubit state $\ket{\Psi_{\rm init}}$ differs from either the $\ket{0}$ or $\ket{1}$ qubit levels sketched in Fig.~\ref{fig:fusion}(b).  A basis change (related to so-called F-symbols in the underlying topological quantum field theory) allows one to recast $\ket{\Psi_{\rm init}}$ in terms of the latter, yielding the equal superposition
\begin{equation}
    \ket{\Psi_{\rm init}} = \frac{1}{\sqrt{2}}(\ket{0} + \ket{1}).
\end{equation}
(See, e.g., Ref.~\onlinecite{Aasen2016} for details and conventions leading to the above relation.)  That is, the qubit has been initialized into a state for which the left pair and the right pair of Ising anyons each fuse into the $I$ and $\psi$ channels with equal probability.   
After the attempted creation of the second Ising-anyon pair, interferometric readout of the anyon charge on, say, the right set of holes returns the following possible outcomes, along with their probabilities:
\begin{itemize}
    \item a single $\sigma$ particle with probability $1-p_{\rm suc}$ (i.e., the attempt failed),
    \item an $I$ particle with probability $p_{\rm suc}/2$,
    \item a $\psi$ particle with probability $p_{\rm suc}/2$.
\end{itemize}
Resolving equal probabilities for the latter two outcomes via successive initialization protocols would yield compelling evidence for the fusion rule $\sigma \times \sigma = I + \psi$ as claimed.  

A skeptic may reasonably counter, however, that equal probabilities for $I$ and $\psi$ could alternatively arise for more trivial reasons, e.g., from processes that randomly split a pair of $\psi$ fermions into the left and right halves of the qubit.  The single-$\sigma$ outcome that occurs with probability $1-p_{\rm suc}$ would be unaffected by such events, since $\psi \times \sigma = \sigma$.  Moreover, the $I$ and $\psi$ outcomes could still occur with equal probability.  A relatively simple control experiment can plausibly rule out this undesirable scenario: Run the above fusion experiment for the case where the first attempted Ising anyon pair creation failed, such that the left and right halves begin in the $I$ sector rather than the $\sigma$ sector.  When the second Ising anyon pair creation \emph{also} fails, interferometric detection should preferentially return $I$.  By contrast, random $\psi$ generation would be expected to return $I$ and $\psi$ again with equal likelihood, even though this control experiment has nothing to do with nontrivial fusion rules.

For the electrical approach, the nontrivial fusion rule $\sigma \times \sigma = I + \psi$ can be similarly demonstrated by initializing the qubit via the steps shown in Figs.~\ref{fig:electric_fusion}(a), and then interferometrically reading out the resulting state [see Fig.~\ref{fig:electric_qubit}(c)]. Given that the probability of success for anyon-pair creation is $p_{\rm suc} = 1$ in the electrical approach, this readout measurement should return outcomes $I$ and $\psi$ with equal $50\%$ probabilities, thus reflecting the nontrivial fusion rule above. For a control experiment ruling out more trivial explanations, one can initialize the qubit by following the steps in Fig.~\ref{fig:electric_fusion}(b) instead of Fig.~\ref{fig:electric_fusion}(a), in which case the final readout measurement should return the same outcome $I$ every single time.

\begin{figure*}[ht]
    \centering
    \includegraphics[width=\linewidth]{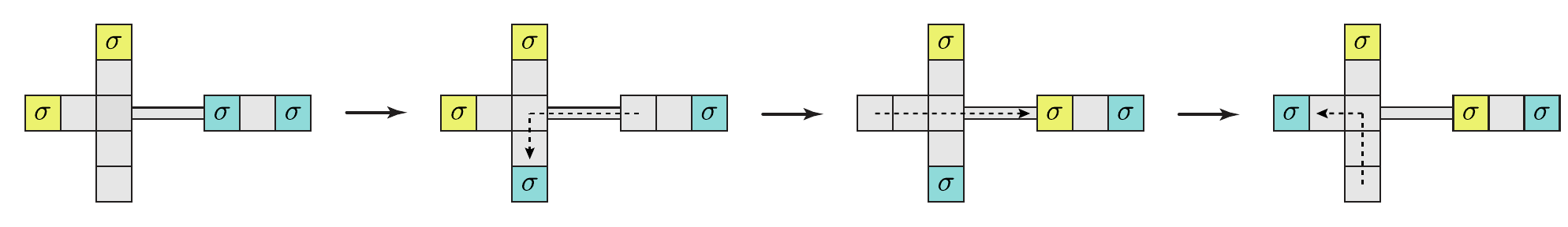}
    \caption{
    \textbf{Braiding experiment in the magnetic tunnel junction setup.}
    The qubit is initialized in the logical state $\ket{0}$ such that Ising-anyon pairs of the same color fuse to the identity channel $I$.
    Braiding is facilitated by four additional tunnel junctions which allow the anyons to be moved past one another by the sequence of inchworm moves depicted.
    Here, exchange of the first and third anyons enacts a unitary rotation gate $\rotGate{Y}{-\pi/2}$ on the qubit.
    Interferometric readout of the resulting qubit state allows for verification of the non-Abelian braiding statistics underpinning the logical gate.
    }
    \label{fig:braiding_tj}
\end{figure*}

\begin{figure*}[t]
    \centering
    \includegraphics[width=\linewidth]{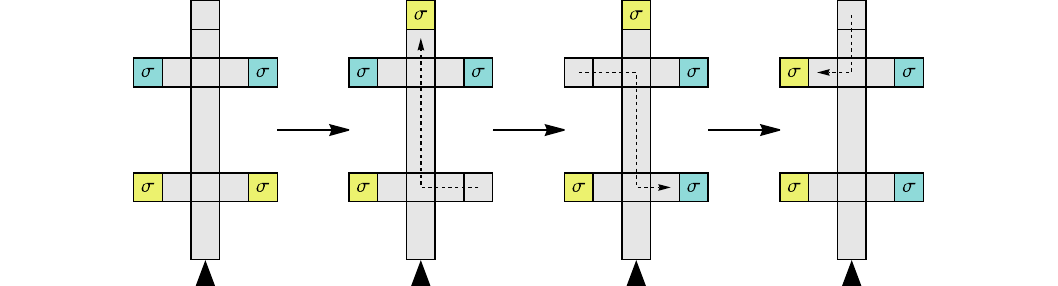}
    \caption{\textbf{Braiding experiment in the electrical approach.} Following qubit initialization in state $\ket{0}$ [see Fig.~\ref{fig:electric_fusion}(b)], two Ising anyons $\sigma$ are exchanged with each other via appropriate inchworm moves. (Note that each of the blue and yellow pairs fuse to the identity channel $I$.) Repeatedly implementing this exchange protocol both once and twice after qubit initialization, one can statistically demonstrate non-Abelian braiding of Ising anyons, as described in the text.}
    \label{fig:electric_braiding}
\end{figure*}

\subsection{Lifetime demonstration}\label{ss:lifetime}

Various factors influence the coherence times of a topological qubit~\cite{Aasen2016, Hell_2016, Hu_2015, Schmidt_2012, Rainis_2012, Goldstein_2011}.  Finite-temperature and/or non-equilibrium effects can generate bulk anyons that surreptitiously braid around a subset of anyons encoding the qubit---producing dephasing errors.  Alternatively, such effects can generate bulk emergent fermions that decay in a way that flips the qubit or even exits the computational subspace entirely.  (One can view the latter mechanism as analogous to quasiparticle poisoning events in topological-superconductor-based qubit designs~\cite{Aasen2016, Hassler_2011}.)  We will assume in what follows that bulk quasiparticles are \emph{not} generated on experimentally relevant time scales, and that the system instead remains confined to the low-energy qubit subspace with high probability.  This situation can arise provided that the temperature and the frequency associated with environmental noise both fall well below the spin liquid's bulk excitation gap.  

Even in this scenario, a topological qubit will still generically be subject to decoherence.  To see why, it is useful to invoke a further idealization by assuming that the topological qubit is encoded in exactly degenerate states that can not be distinguished via any local measurements and hence is perfectly immune to arbitrary local noise sources.  Exact degeneracy means that the qubit's oscillation frequency $\omega_0$, given by the \emph{time-averaged} splitting between the qubit $\ket{0}$ and $\ket{1}$ states, vanishes.  Perfect noise immunity further implies that the qubit's \emph{instantaneous} splitting $\omega(t)$ is pinned to its time-averaged splitting, i.e., noise can not stochastically modulate the $\ket{0}$ and $\ket{1}$ energy difference; the dephasing time $T_2$ correspondingly diverges in this ideal limit.  

In practice, the degeneracy will not vanish exactly, yet can generically be exponentially small in the spatial separation between Ising anyons encoding the qubit.  Here the qubit states exhibit neither perfect local indistinguishability nor perfect immunity against local noise.  One could, for instance, locally measure the difference in energy density between $\ket{0}$ and $\ket{1}$, with the measurement becoming increasingly difficult as system parameters are tuned to decrease the exponentially small splitting towards the perfect-qubit limit.  
As a corollary, imperfect noise immunity implies that local noise can also (weakly) modulate the $\ket{0}$ and $\ket{1}$ energy difference, yielding a finite $T_2$ time that tends to increase and eventually diverge on approaching the perfect-qubit limit.  

The link between qubit energy splitting, local indistinguishability, and dephasing implies scaling relations between $\omega_0$ and $T_2$ that are unique to topological qubits \cite{Aasen2016}.  For a particular noise model, Ref.~\onlinecite{Aasen2016} in particular derived the relation
\begin{equation}
    \omega_0 \sim \frac{1}{T_2} \sim e^{-L/\xi},
    \label{T2relation}
\end{equation}
where $L$ is the separation between Ising anyons that dominate the residual energy splitting and $\xi$ is the spin liquid correlation length.  (Relations linking the qubit relaxation time $T_1$ with $T_2$ also arise \cite{Aasen2016}.) Verifying Eq.~\eqref{T2relation}---which crucially links time-averaged behavior ($\omega_0$) to fluctuations ($T_2$)---would validate the qubit's predicted ability to encode quantum information in an intrinsically fault-tolerant manner.  

Fortunately, the same architectures used for fusion enable testing this prediction using a standard Ramsey protocol: (i) Initialize the qubit into $\ket{0}$, (ii) apply a $\pi/2$ pulse to rotate the qubit to an equal superposition of $\ket{0}$ and $\ket{1}$, (iii) wait for a time $t_{\rm wait}$, (iv) apply a second $\pi/2$ pulse, and then (v) read out the qubit state in the computational basis.  The $t_{\rm wait}$ dependence of the readout probabilities reveals both $\omega_0$ and $T_2$.  
Requisite $\pi/2$ pulses can be implemented using unprotected gates, e.g., by pulsing the central bridge in Fig.~\ref{fig:fusion} to couple the inner Ising anyons for a prescribed duration; see Sec.~\ref{PhaseGate} for further discussion.  
(One can alternatively generate protected $\pi/2$ pulses in more elaborate geometries---considered later---that enable braiding the inner Ising anyons.)
Moreover, altering the ratio $L/\xi$ on the right side of Eq.~\eqref{T2relation} may be achieved in a fixed device geometry by changing parameters such as the magnetic field to vary $\xi$ at fixed $L$.

\subsection{Non-Abelian braiding protocols}

Although the minimal architectures in Figs.~\ref{fig:fusion}(a) and \ref{fig:electric_qubit}(a) are sufficient for encoding, initializing, and reading out a single topological qubit, they do not permit the non-Abelian braiding processes that would perform topological quantum gates on the qubit. We can remedy this deficit at the expense of adding more tunable regions, as shown in Figs.~\ref{fig:braiding_tj} and \ref{fig:electric_braiding} for the two approaches.

With these expanded architectures, non-Abelian braiding can be demonstrated in each approach by initializing the qubit in the state $\ket{0}$ and then exchanging two anyons that do not fuse to the identity channel $I$ through appropriate inchworm moves (see Figs.~\ref{fig:braiding_tj} and \ref{fig:electric_braiding}). Since the qubit ends up in an equal superposition of $\ket{0}$ and $\ket{1}$, interferometric readout is expected to return outcomes $I$ and $\psi$ with equal $50\%$ probabilities, demonstrating that an exchange of two anyons corresponds to a nontrivial quantum gate. A simple control experiment can rule out more mundane explanations for this maximally random distribution---e.g., random $\psi$ generation addressed also for the fusion protocols above.  Specifically, one may repeat the steps in Figs.~\ref{fig:braiding_tj} and \ref{fig:electric_braiding} \emph{twice} after each successful initialization of the qubit. This double exchange rotates the state from $\ket{0}$ to $\ket{1}$, and interferometric readout should then deterministically return outcome $\psi$.

\section{Demonstration of elementary quantum gates}\label{sec:gates}

Having described how to nucleate, braid, and fuse anyons, we now consider how to implement the elementary quantum gates which are necessary for universal quantum computation.
A standard choice of a universal gate set is the set of Clifford unitaries supplemented by a non-Clifford ``magic'' gate.
Recall that the Clifford group is generated by (i) the single-qubit phase gate $P = e^{i\pi Z/4}$, (ii) the basis-changing Hadamard gate $H = (X + Z)/\sqrt{2}$, and (iii) the two-qubit controlled-NOT gate $\textrm{CNOT} = e^{i\frac{\pi}{4}(1-Z)\otimes(1-X)}$.
In the following, we adopt the notation $R_O(\theta) \equiv e^{i\theta O / 2}$ as the unitary rotation gate through angle $\theta$ with respect to Pauli operator $O$ acting on the logical qubits.

We focus on minimal demonstrations of one- and two-qubit gates.
For the multi-qubit scenario, the implementation of logical operations depends sensitively on the choice of encoding.
A set of $2n$ Ising anyons forms a $2^{n-1}$ dimensional ``fusion space'' in which one can encode quantum information ($n = 2$ corresponds to the single-qubit case discussed in the previous section).  
A logical encoding corresponds to an embedding that maps the logical qubit codespace to a subspace of the fusion space.
In the following, we consider two encodings of $n_q$ logical qubits:
(i) the \emph{sparse} encoding, using $4n_q$ anyons, and
(ii) the \emph{dense} encoding, using $2n_q+2$ anyons.  Encoding (i) straightforwardly extends the single topological qubit by using an Ising anyon quartet for each additional qubit, only partially taking advantage of the degenerate subspace generated by the anyons.  Encoding (ii) fully exploits the accessible states by using only \emph{two} Ising anyons for each additional qubit.  Despite the latter's more economical use of Hilbert space, however, encoding (i) is more commonly pursued because encoding (ii) is more susceptible to corruption by errors~\cite{Sarma_2015}.  

\begin{figure}
    \centering
    \includegraphics[width=\columnwidth]{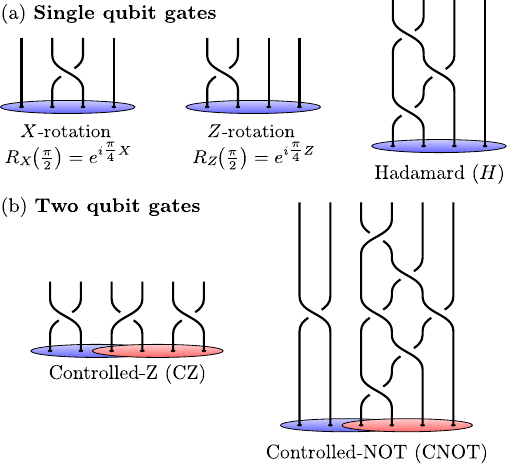}
    \caption{
    \textbf{Elementary quantum gates via braiding.}
    (a) Single-qubit braid gates for a logical qubit encoded in 4 Ising anyons.
    All such single-qubit braid gates may be implemented in the minimal architectures of Figs.~\ref{fig:braiding_tj} and \ref{fig:electric_braiding}.
    (b) Two-qubit braid gates for logical qubits encoded in 6 Ising anyons. The two qubits correspond to the two groups of four anyons inside the blue and red ellipses.
    }
    \label{fig:braid_gates}
\end{figure}

\subsection{Topologically protected braid gates}

Since the logical state is encoded in the fusion channels of bulk anyons, braiding the anyons applies a topologically protected operation on the encoded state.
Let us denote by $\braid_i$ the braid group generator corresponding to a clockwise exchange of anyons $\sigma_i$ and $\sigma_{i+1}$.
For $2n$ anyons, these generators are faithfully represented by the Clifford algebra representation of the group $SO(2n)$~\cite{Georgiev_2009, Nayak_1996}.
We may thus identify the anyon fusion space with the parity-even sector of the Hilbert space for $n$ qubits.
Let $\gamma_j$ be the emergent Majorana zero mode operator bound to $\sigma_j$, and let $f_j = (\gamma_{2j-1} + i\gamma_{2j})/2$ span a corresponding set of $n$ complex fermions that one can use to conveniently enumerate the degenerate manifold.
Then the braid operators may be written, up to a phase, as Clifford unitaries acting on the qubits,
\[
    \braid_{j} = e^{\frac{\pi}{4}\gamma_{j}\gamma_{j+1}}.
\]

For a single logical qubit, there is no difference between the sparse and dense encodings, both of which use 4 anyons.
Here, braiding generates the full set of Clifford operations~\cite{Georgiev_2006, Ahlbrecht2009Mar, Georgiev_2008}.
Nearest-neighbor braids yield the $\pi/2$ rotation gates $\braid_1 \sim \rotGate{Z}{\pi/2}$ and $\braid_2 \sim \rotGate{X}{\pi/2}$ acting on the logical qubit.
Composition of such braids give a simple implementation of the 
Hadamard gate, as depicted in Fig.~\ref{fig:braid_gates}(a).
In Figs.~\ref{fig:braiding_tj} and \ref{fig:electric_braiding}, we show minimal architectures for demonstrating single-qubit braid gates for the tunnel-junction and charge-based schemes, respectively.

With two logical qubits, the sparse and dense encodings yield distinct sets of logical gates generated via braiding.
The sparse encoding gives a straightforward generalization of the single-qubit case, using four anyons per qubit.
As such, single-qubit gates remain the same as before, involving only the group of four anyons corresponding to the qubit being acted upon.
Since the total topological charge of each qubit is trivial, the two-qubit SWAP gate $\tfrac12(1 + X_1X_2 + Y_1Y_2 + Z_1Z_2)$ may be implemented by a sequence of braids which exchange the four anyons encoding each of the qubits.
More generally, however, braiding in the sparse encoding may only produce unentangled (separable) two-qubit states~\cite{Bravyi_2005}.
This limitation can be seen from the fact that non-trivial entangling gates [e.g., controlled-NOT (CNOT) or controlled-Z (CZ)] require access to the joint parity $Z_1Z_2$ of the logical qubit, which corresponds to a 4-fermion unitary not possible via braiding alone.

By contrast, the dense encoding of two logical qubits employs only 6 anyons.
This is achieved by allowing the two qubits to ``share'' two anyons such that the following correspondence between logical operators and fusion channels holds: $Z_1 \sim (2f_1^\dagger f_1 - 1)$, $Z_2 \sim (2f_3^\dagger f_3 - 1)$, and $Z_1 Z_2 \sim (2f_2^\dagger f_2 - 1)$.
To this end, single-qubit gates remain largely the same from the single-qubit case, with $\rotGate{Z_1}{\pi/2} \sim \braid_1 = e^{\frac{\pi}{4}\gamma_1\gamma_2}$ and $\rotGate{Z_2}{\pi/2} \sim \braid_5$.
As the joint parity $Z_1 Z_2$ now corresponds to a fermion bilinear, two-qubit entangling gates may be implemented via anyon braiding~\cite{Georgiev_2006}.
Figure~\ref{fig:braid_gates}(b) depicts the braid sequence that implements CZ and  CNOT gates on two densely encoded qubits.
Thus, we may implement the full two-qubit Clifford group for 6 anyons exclusively by braiding.
Crucially, this statement does \emph{not} generalize to the case of more than two densely encoded qubits.
There we no longer have access simultaneously to individual qubit Clifford gates and pairwise two-qubit joint parities $Z_j Z_{j+1}$.
Moreover, it becomes nontrivial to implement the SWAP gate since naive exchange of the pairs of anyons corresponding to two qubits implements the fermionic fSWAP gate.
Nonetheless, six anyons suffice to demonstrate topologically protected two-qubit Clifford gates and explore coherence times therein.

\subsection{Measurement-assisted gates}

\begin{figure}[t]
    \centering
    \includegraphics[width=\columnwidth]{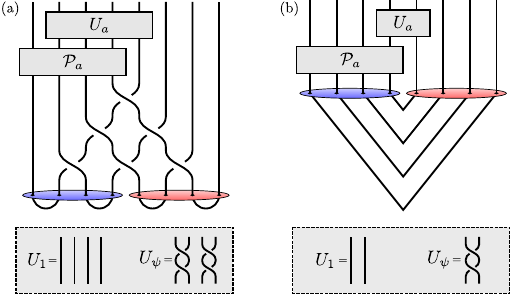}
    \caption{\textbf{Measurement-assisted Bell pair generation}.
    Two different schemes for preparing Bell pairs from two logical qubits (blue and red ellipses) encoded in 4 Ising anyons each.
    Both schemes employ projective readout $\mathcal{P}_a$ of the topological charge $a$ for a group of 4 anyons followed by a unitary correction $U_a$ conditioned on the measurement outcome.
    When the measurement reveals the topological charge to be in the vacuum sector ($a=1$), we have successfully prepared the logical Bell pair $\tfrac{1}{\sqrt{2}}(\ket{00} + \ket{11})$.
    If instead we measure a fermionic topological charge ($a=\psi$), the state lies outside the logical codespace and must be corrected by additional braiding.
    (a) If the two logical qubits are initially prepared in the logical $\ket{0}$ state, we follow the approach of Ref.~\onlinecite{Bravyi_2006}, applying a sequence of braids prior to measuring the topological charge.
    (b) If pairs of anyons are nucleated from vacuum and dragged apart so as to give a nested structure, no additional braids are necessary prior to measurement.
    Moreover, the correction after observing $a=\psi$ needs fewer braids than in (a).
    The need for braiding can be fully eliminated by instead resetting the state via measuring the total charge of the central 4 anyons, then repeating until we get the desired measurement outcome.
    }
    \label{fig:Bell_Pair}
\end{figure}

In both qubit architectures we consider here, anyon interferometry allows for measurement of the total topological charge within an enclosed region of the spin liquid.
Such measurements not only enable quantum-state readout, but also enrich the realizable set of logical operations.
Namely, supplementing anyon braiding with topological charge measurements allows for the full Clifford group to be implemented for an arbitrary number of logical qubits.

As a demonstration of this measurement-assisted gate set, consider the sparse encoding of two qubits in the fusion space of 8 anyons.
Preparing a maximally entangled Bell state $\frac{1}{\sqrt{2}}(\ket{00} + \ket{11})$ requires a four-fermion unitary operation which cannot be realized by braiding.
However, measuring the total charge of anyons $\sigma_{3,4,5,6}$ projects the state into an eigenspace of the joint parity operator $Z_1 Z_2$.
Thus, an EPR pair may be produced by a sequence of braids and measurements,
\begin{gather*}
    \frac{1}{\sqrt{2}}\left(\ket{00} + \ket{11}\right) = U_\pm P_{ZZ}^\pm H_1 H_2 \ket{00},\\
    U_+ = \mathds{1}, \quad U_- = X_2,
\end{gather*}
where $P_{ZZ}^\pm = \frac12(1 \pm Z_1 Z_2)$ are projectors onto the parity sectors and $U_\pm$ is a unitary correction conditioned on the measurement outcome.
If the anyon pairs may be nucleated deterministically (as in the charge-based scheme), we may slightly simplify this protocol by nucleating four nested pairs of Ising anyons rather than using Hadamard gates [see Fig.~\ref{fig:Bell_Pair}(b)].
More generally, the entangling CZ gate may be applied to an arbitrary two-qubit input state by a combination of measurements and conditional braid gates, with the assistance of an ancillary pair of anyons.

\subsection{Non-topological phase gates}\label{sec:phase_gates}

Realizing a \emph{universal} gate set in this setting necessarily requires that we supplement topological braid operations with a non-topological operation for non-Clifford gates or magic state injection.
To this end, we consider here how to demonstrate generic single-qubit rotation gates.
Broadly, there are two approaches to implementing generic phase gates:
(i) temporarily inducing a level splitting so as to acquire a dynamical phase, or
(ii) using anyon interferometry to effectively achieve a superposition of braid operations~\cite{Bonderson_2010, Clarke_2010}.
Here we describe minimal setups required for demonstrating non-topological phase gates by both of these approaches.

\subsubsection{Interaction-based phase gate}
\label{PhaseGate}

As discussed in Sec.~\ref{ss:lifetime}, the two logical qubit states are degenerate when the anyons are sufficiently well separated so as to suppress interactions.
However, interactions may in principle be harnessed for non-topological operations~\cite{Bravyi_2005}.
Suppose that we allow two anyons $\sigma_i$ and $\sigma_j$ to hybridize for a prescribed time $\Delta t$.
The interactions induce a finite energy splitting $\omega$ in the basis where $\sigma_i$ and $\sigma_j$ have definite fusion channel.
The net effect is then to apply a unitary gate $\rotGate{\hat{O}}{\theta=\omega \Delta t}$, where the operator $\hat{O} = Z, X$ depends on which anyon pair one couples.
A minimal demonstration of this unprotected phase gate can be achieved in the single-qubit architectures depicted in Figs.~\ref{fig:fusion} and \ref{fig:electric_qubit} by first initializing the qubit in a computational basis state, then inducing interactions for a prescribed time, and finally verifying the final qubit state via interferometric readout.
In the tunnel junction setup, we can allow adjacent anyons to hybridize by temporarily tuning the intervening junction(s) to a trivial state.
(Since the non-Abelian anyons in this approach are only metastable, the hybridization must occur on a time scale shorter than that at which the anyons relax.)  
Interactions may be similarly generated in the electrical scheme by bringing the anyons close together via a sequence of ``inchworm'' moves as described previously.
This method requires no additional machinery beyond what was already required for qubit initialization and readout, offering a simple proof of principle that can be used for qubit lifetime demonstrations as discussed in Sec.~\ref{ss:lifetime}.
Nonetheless, we note that if the level splitting $\omega$ varies or oscillates rapidly with respect to the tuning parameter (e.g., anyon separation~\cite{Cheng_2009}), it may be challenging to accurately apply a desired phase gate.

\subsubsection{Interferometry-based phase gates}

\begin{figure}[!t]
    \centering
    \includegraphics[width=\columnwidth]{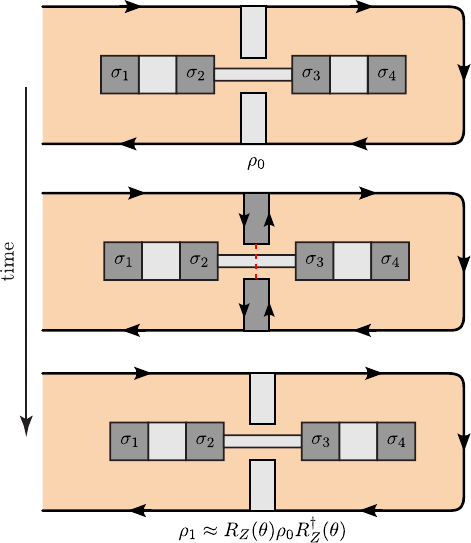}
    \caption{
    \textbf{Tunneling-based phase gate scheme in the magnetic tunnel junction setup.}
    Starting from an initial logical-state density matrix $\rho_0$ encoded in the Ising anyons $\sigma_{1,2,3,4}$, the edge is deformed to produce a point contact with Ising-anyon tunneling (dashed red line) for a finite time $\Delta t$.
    The resulting density matrix $\rho_1$ approximately corresponds to $\rho_0$ acted upon by a phase gate $R_Z(\theta)$, where the angle $\theta$ depends on the tunneling strength, the time $\Delta t$, and the edge temperature $T$ through Eq.~\eqref{eq:tunneling_phase_angle}.
    }
    \label{fig:phase_gate_2}
\end{figure}

\begin{figure}[!t]
    \centering
    \includegraphics[width=\columnwidth]{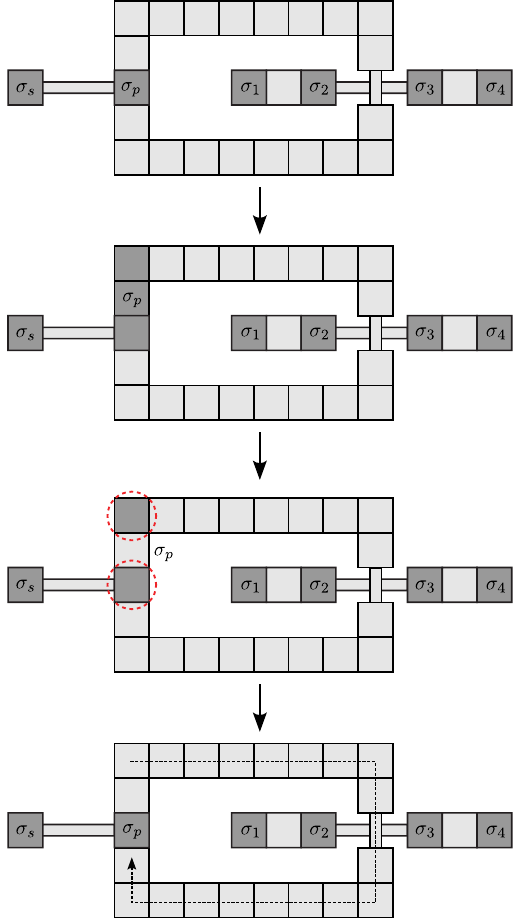}
    \caption{
    \textbf{Discrete-time phase gate scheme in the magnetic tunnel junction setup.}
    The procedure starts with a logical qubit encoded in the Ising anyons $\sigma_{1,2,3,4}$, as well as two auxiliary Ising anyons $\sigma_s$ (spectator) and $\sigma_p$ (probe).
    The hole on which $\sigma_p$ resides is initially enlarged and then smoothly pinched in half, resulting in a superposition of two possible anyon locations (dashed red circles).
    To produce the necessary interference, one of the two holes on which $\sigma_p$ may be residing is dragged around qubit anyons $\sigma_1$ and $\sigma_2$, resulting in a braid phase that depends on the fusion channel of $\sigma_1$ and $\sigma_2$.
    Finally, the two holes for $\sigma_p$ are merged.
    }
    \label{fig:phase_gate_3}
\end{figure}

Whereas the interaction-based phase gate distinguishes between the two qubit states by an induced energy splitting, the interferometry-based scheme relies instead upon anyon braiding statistics.
Consider a single logical qubit encoded in anyons $\sigma_i$ for $i=1,2,3,4$ (enumerated from left to right as in Figs.~\ref{fig:phase_gate_2} and \ref{fig:phase_gate_3}).
We allow an Ising anyon $\sigma_\textrm{probe}$ to travel along two interfering paths which enclose qubit anyons $\sigma_i$ and $\sigma_j$, resulting in a braid phase of $+1$ or $-1$ for fusion channels $I$ and $\psi$, respectively.
This process enacts a logical phase gate whose angle $\theta$ depends on microscopic parameters setting (i) the relative dynamical (non-braid) phase for the two paths, (ii) the relative amplitudes for the two paths, and (iii) the number of anyons we send through the interferometer.
We distinguish two types of interferometry-based phase gates: continuous-time and discrete-time gates.

%%% Continuous-time gates

For both the magnetic tunnel junction and electrically gated setups, we have already demonstrated how to perform anyon interferometry for qubit readout.
The same architecture can be used to implement arbitrary phase gates.
With tunnel junctions, consider the process depicted in Fig.~\ref{fig:phase_gate_2}.
We allow for Ising anyon tunneling at a pinch point for finite time $\Delta t$.
Since anyons $\sigma_3$ and $\sigma_4$ are enclosed by the interfering paths, the tunneling enacts a phase gate $\rotGate{Z}{\theta}$.
For fixed geometry and temperature, the gate angle $\theta$ is tunable via the time $\Delta t$ and tunneling strength $t_\sigma$, which depends exponentially on the distance across the point contact~\cite{Fendley_2009, Chen_2009, Cheng_2009}.
At finite temperature, this tunneling process also induces finite decoherence of the logical qubit state.
See Appendix~\ref{app:continuous_phase_gate} for further details on how the gate angle $\theta$ and the decoherence rate depend on the relevant tuning parameters.
In the electrical setup, a similar phase gate may be implemented.
Passing a current from left to right through the interferometry loop depicted in Fig.~\ref{fig:electric_qubit}(c) drags electron-anyon pairs along one of two paths enclosing the qubit anyons.
In this case, there are fewer tuning parameters since the interferometry no longer relies on edge tunneling of Ising anyons.
Furthermore, since the anyons are bound to electrons in the semiconductor, interference is now sensitive not only to the anyonic braiding statistics but also to the Aharonov-Bohm phase~\cite{Halasz_2024}.
As such, the phase-gate angle $\theta$ may controlled by tuning the relative amplitudes for the two paths and the Aharonov-Bohm phase (through, e.g., an out-of-plane magnetic field).

%%% Discrete-time gates

\begin{figure*}[t]
   \centering
  \includegraphics[width=\linewidth]{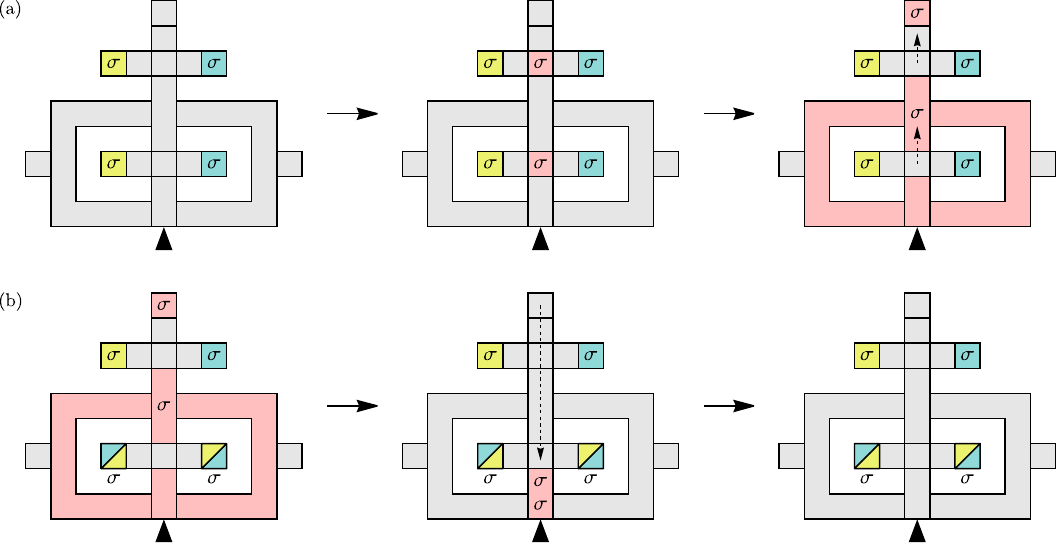}
   \caption{\textbf{Discrete-time phase gate scheme in the electrical approach.} (a) First step at zero time ($t=0$). The qubit is initialized in the superposition state $(\ket{0} + \ket{1}) / \sqrt{2}$ [see Fig.~\ref{fig:electric_fusion}(a)]; an auxiliary pair of Ising anyons $\sigma$ is created (see Fig.~\ref{fig:electric_initialization}); one auxiliary anyon is delocalized among the four islands that form a ring around the bottom pair of qubit anyons. (b) Second step at finite time ($t>0$). The auxiliary Ising anyons are brought together and annihilated. Due to the time evolution beforehand, the qubit ends up in the new state $(\ket{0} + e^{i \theta} \ket{1}) / \sqrt{2}$, where the phase-gate angle $\theta$ is proportional to the evolution time $t$.}
   \label{fig:phase_gate_electric}
\end{figure*}

Rather than relying upon a continuous stream of interfering anyons, we may instead use discrete-time operations to perform a tunable phase gate with individual probe anyons.
In the tunnel junction architecture, interferometric readout and the corresponding phase gate rely upon deforming the spin liquid edge to form a pinch point where Ising anyon tunneling takes place.
To interfere with individual anyons, we must instead move the whole process into the bulk of the spin liquid.
As depicted in Fig~\ref{fig:phase_gate_3}, we may nucleate a pair of Ising anyons consisting of a spectator $\sigma_s$ and a probe $\sigma_p$.
In order to get two interfering paths for $\sigma_p$, we first grow the hole on which it resides and then pinch it off, resulting in a superposition between $\sigma_p$ residing on one of two holes.
One of these two holes may then be dragged around two qubit anyons to acquire the braid phase before merging back to the other hole.
The relative amplitude of the two interfering paths can be tuned by varying the speed and position at which the hole is pinched off.
Moreover, by modulating the circumference $L$ of one of the two holes after pinching, we may induce an energy difference (recall $E \sim 2\pi / L$) and thus a dynamical phase between the interfering paths.
In comparison to the tunneling-based phase gate, this refined scheme requires a more complicated architecture and more movement of anyons in the bulk.
However, this additional effort is rewarded with decoherence effects being better controlled here.
In Appendix~\ref{app:discrete_phase_gate}, we provide further details on the effective logical operation applied to the qubit by this process.

An analogous discrete-time phase gate may also be implemented in the electrically gated setup with minimal modification to the single-qubit architecture already introduced above.
Specifically, as depicted in Fig.~\ref{fig:phase_gate_electric}, one nucleates a pair of auxiliary anyons from the vacuum, delocalizes one of these auxiliary anyons along the interferometry loop surrounding a pair of qubit anyons, and then annihilates the two auxiliary anyons back into the vacuum after a finite time $t$.
Since the kinetic energy of the auxiliary anyon along the interferometry loop depends on the fusion channel of the two qubit anyons inside~\cite{Halasz_2024}, this operation implements an appropriate phase gate of angle $\theta$.
Importantly, the energy difference between the two fusion channels can be as little as $\qty{0.1}{\micro\eV}$ for the realistic parameters described in Ref.~\onlinecite{Halasz_2024}, which means that the phase-gate angle can be controlled with an accuracy $\theta \sim 0.1$ on nanosecond timescales $t$.
In addition to varying the operation time $t$, the phase-gate angle $\theta$ may also be tuned by an out-of-plane magnetic field via the Aharonov-Bohm effect.

\subsubsection{Magic-state distillation for imperfect phase gates}

Thus far, we have outlined several distinct approaches for implementing non-Clifford phase gates.
All of these methods are susceptible to noise or imperfect calibration, which will tend to spoil the encoded quantum state and its evolution.
Nonetheless, noisy phase gates in tandem with ideal (topologically protected) Clifford operations allow one to realize clean phase gates via magic-state distillation~\cite{Bravyi_2005}.
In particular, given several noisy copies of a magic state, an appropriate sequence of measurements and unitary gates may be used to produce a single higher fidelity magic state.
While `magic state distillation' will require many more qubits and gates or tunnel junctions, it is a key operation to demonstrate when scaling beyond one or two qubits.
In the meantime, however, noisy phase gates empower a more robust characterization of the minimal qubit designs presented here.

%%%%%%%%%%%%%%%%%%%

\section{Discussion}
\label{Discussion}

Quantum spin liquids have persisted as a topic of great theoretical and experimental interest for decades, yet the question `What does a topological qubit based on non-Abelian spin liquids look like?'~has remained largely unanswered.  
Building on recent theory insights, we have proposed two classes of potentially scalable topological qubit designs for non-Abelian spin liquids realized in Kitaev materials---and potentially other families of magnetic insulators.
The first class uses Kitaev materials integrated into magnetic tunnel junction arrays that enable dynamically tuning select patches of the system between spin-liquid and trivial phases. The second class couples Kitaev materials to semiconductors---binding the non-Abelian anyons to electric charges and enabling all-electrical control of the charge-anyon composites.  

\begin{figure}[t]
    \centering
    \includegraphics[width=\linewidth]{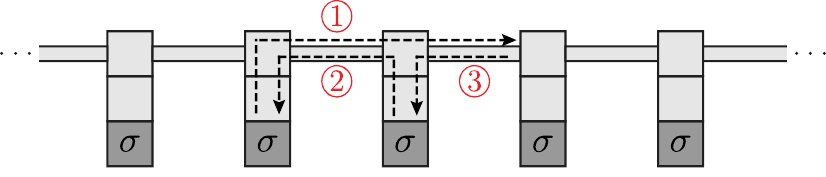}
    \caption{
    \textbf{Scalable qubit architecture in the magnetic tunnel junction setup.}
    Generalizing the single-qubit braiding scheme from Fig.~\ref{fig:braiding_tj}, this architecture consists of a series of storage points where Ising anyons are held between braiding operations.
    These storage points are connected by horizontal bridges that are used not only for nucleating Ising-anyon pairs but also for lateral movement of anyons past one another.
    The elementary braiding operation is implemented by a sequence of moves (dashed lines) making use of this auxiliary space.
    }
    \label{fig:scalable_architecture}
\end{figure}

Both designs offer pathways to anyon creation, manipulation, and interferometric readout desired for fault-tolerant quantum information applications.  We have specifically identified architectures and protocols that one can exploit to experimentally demonstrate non-Abelian fusion rules and braiding properties; extract topological qubit lifetimes and oscillation frequencies to verify topological protection of quantum information; and establish a gate set needed for universal quantum computation in this setting.   Our emphasis was on minimal devices needed for establishing each of these fundamental elements of topological quantum computation with spin liquids.  Nevertheless, the basic design principles naturally extend to more complex architectures.  For instance, Fig.~\ref{fig:scalable_architecture} presents a multi-qubit array that allows for braiding any pair of non-Abelian anyons.

Even the simplest single-qubit devices require nontrivial experimental advances to realize in practice.  Our work does, however, spotlight nearer-term challenges and opportunities: 
Can one optimize the probabilistic anyon-generation protocol in the magnetic tunnel junction design to increase the success probability towards unity?  Are there variations on the magnetic tunnel junction approach that provide local control of the Kitaev material with less experimental overhead?  For example, one might attempt to reposition the pair of ferromagnetic metals in Fig.~\ref{fig:generation}, instead  placing both side-by-side \emph{above} the Kitaev material \footnote{We thank Charlie Marcus for suggesting this possibility.}.  In the parallel configuration, the adjacent part of the Kitaev material clearly feels a net exchange field; in the antiparallel configuration, the net exchange field effectively \emph{spatially} averages to zero provided the spin-liquid correlation length is long compared to the size of the ferromagnets.  Such a modification could simplify fabrication while also being applicable to bulk Kitaev materials.  In the electrical design, can first-principles calculations provide useful guidance for designing appropriate Kitaev material-semiconductor interfaces?  More broadly, the electrical approach naturally applies also to \emph{Abelian} spin liquids, where charge-Abelian anyon bound states can form via a similar mechanism.  Can one utilize that technique to control topological qubits based on Abelian spin liquids hosting extrinsic non-Abelian defects~\cite{Bombin}?

\section*{Acknowledgments}

We are grateful to Arnab Banerjee, Charlie Marcus, and Alan Tennant for insightful discussions. 
This work was primarily supported by the U.S.~Department of Energy, Office of Science, National Quantum Information Science Research Centers, Quantum Science Center. 
 Additional support was provided by the Caltech Institute for Quantum Information and Matter, an NSF Physics Frontiers Center with support of the Gordon and Betty Moore Foundation through Grant GBMF1250.
KK was additionally supported by an NSF Graduate Fellowship under Grant No.~DGE 2146752

\bibliography{notesbib}

\clearpage
%%%%%%%%%%%%%%%%%%%%%%%%%%%%%%%%%%%%%%%%%%%%%%%%%%%%%%%%%%%%%%%%%%%%%%%%%%%%
\begin{appendix}

\section{Interferometry-based phase gates}\label{app:interferometric_phase_gates}

In Sec.~\ref{sec:phase_gates} of the main text we laid out several approaches for implementing arbitrary (non-Clifford) single-qubit phase gates using anyon interferometry.
Here we provide additional details on how the gate angle $\theta$ depends on microscopic tuning parameters, with consideration of the typical parameters expected for candidate material \alpRuCl{3}.
Furthermore, we discuss sources of decoherence and errors, as well as mitigation strategies therein.
In Sec.~\ref{app:continuous_phase_gate}, we focus on the continuous-time scheme in the magnetic tunnel junction setup.
Then in Sec.~\ref{app:discrete_phase_gate}, we consider the discrete-time approach more generically, with application to both the tunnel junction and electrically gated setups.

\subsection{Continuous-time scheme with magnetic tunnel junctions}\label{app:continuous_phase_gate}

The continuous-time interferometry phase gate described in Sec.~\ref{sec:phase_gates} was originally introduced in Ref.~\onlinecite{Bonderson_2010} and further refined for the case of neutral anyons in Ref.~\onlinecite{Clarke_2010}.
Here we consider the tunnel junction scheme depicted in Fig.~\ref{fig:phase_gate_2}.
Recall that tunnel junctions are employed to temporarily deform the spin liquid edge, producing a pinch point.
At this point contact we have a finite rate of Ising anyon tunneling, described by the Hamiltonian $H_\textrm{tunnel} = t_\sigma \sigma(x_1)\sigma(x_2) + \textrm{h.c.}$, where $x_{1,2}$ are the coordinates of the pinch point along the edge.
After a time $\Delta t$, the edge is returned to normal, destroying the point contact and thus tuning the tunneling coefficient $t_\sigma$ to zero.
The two interfering paths enclose qubit anyons $\sigma_3$ and $\sigma_4$ (see Fig.~\ref{fig:phase_gate_2}), for which the fusion channel is diagonal in the computational ($Z$) basis of the qubit.
The interferometry process thus yields a final logical state described by the density matrix $\rho_1 \approx R_Z(\theta) \rho_0 R_Z^\dagger(\theta)$.

The phase-gate angle $\theta$ depends on several microscopic parameters.
Let us assume that the spin liquid edge has finite temperature $T_\textrm{edge}$ and velocity $v_\textrm{edge}$.
Defining the dimensionless quantity $\tilde{T} \equiv T_\textrm{edge} \pi \abs{x_2 - x_1} / v_\textrm{edge}$, the interferometry implements a phase-gate with angle
\begin{equation}
    \theta = 4 \Delta t\left(\frac{\lambda \tilde{T}}{\abs{x_2 - x_1}\sinh(\tilde{T})}\right)^{1/8}\textrm{Re}(t_\sigma),
    \label{eq:tunneling_phase_angle}
\end{equation}
where $\lambda$ is a short-range cutoff~\cite{Clarke_2010}.

Coherent tunneling at the point contact generates entanglement between the edge and the bulk anyons.
Tracing out the edge therefore results in decoherence of the logically encoded qubit, with off-diagonal density matrix elements being suppressed by a factor of $e^{-\zeta^2/2}$.
Assuming a separation of scales $\Delta t \gg \abs{x_2 - x_1}/v_\textrm{edge}$ and $\tilde{T} \ll 1$, Ref.~\onlinecite{Clarke_2010} obtained
\begin{equation}
    \zeta^2 = \frac{\abs{x_2 - x_1}}{48 \Delta t v_\textrm{edge}} \theta^2 \tilde{T}^3.
    \label{eq:tunneling_decoherence}
\end{equation}
Observe that both the angle $\theta$ and the decoherence $\zeta^2$ grow linearly with the tunneling time $\Delta t$.
Our aim is then to realize robust control over $\theta$ while minimizing the decoherence rate.
At sufficiently low temperatures $\tilde T \rightarrow 0$, the phase $\theta$ approaches a $\tilde{T}$-independent value
\[
    \theta = 4\Delta t \textrm{Re}(t_\sigma) \left(\frac{\lambda}{\abs{x_2 - x_1}}\right)^{1/8} + \mathcal{O}\left(\tilde{T}^2\right).
\]
If the tunneling strength $t_\sigma$ and lengthscale $\abs{x_2 - x_1}$ are fixed, this allows for straightforward tuning of $\theta$ by means of the tunneling time.
Moreover, in this limit the decoherence rate scales as $T_\textrm{edge}^3$.
Thus by cooling the system sufficiently so that $\tilde{T} \ll 1$, decoherence may be almost entirely suppressed.

Let us now comment on the feasibility of such a phase gate scheme in the context of the Kitaev material \alpRuCl{3}.
We assume an edge velocity $v_\textrm{edge} \sim \qty{e3}{\metre\per\second}$, lattice constant $a \sim \qty{0.6}{\nano\metre}$, and bulk correlation length $\xi_\textrm{bulk}\sim\qty{5}{\nano\metre}$.
The device geometry is constrained by the requirements that the anyons be well separated (relative to $\xi_\textrm{bulk}$) with respect to one another, the spin liquid edge, and the tunnel junctions used to deform the edge. 
Moreover, we require that the path length $\abs{x_2 - x_1}$ not be much larger than the thermal coherence length along the edge $\xi_T \sim v_\textrm{edge}/T_\textrm{edge}$.
To this end, consider fixing $\abs{x_2 - x_1} \sim \qty{100}{\nano\metre}$.
Coherence then requires $T_\textrm{edge} \lesssim \qty{76}{\milli\kelvin}$.
Similarly, achieving $\tilde{T} \ll 1$ requires temperatures $T_\textrm{edge} \lesssim \qty{24}{\milli\kelvin}$.
Note that even for higher temperatures $\tilde{T} \sim 1$, the decoherence given in Eq.~\eqref{eq:tunneling_decoherence} is negligible for gate timescales $\Delta t$ on the order of a nanosecond.
The tunneling strength $t_\sigma$ decays exponentially with the ratio $d/\xi_\textrm{bulk}$, where the distance across the point contact $d$ is fixed by device geometry.
By calibrating the tunneling time $\Delta t$ with respect to experimentally determined $t_\sigma$, the magnetic tunnel junction architecture should enable robust non-topological phase gates with minimal decoherence.
While the requisite temperature $T_\textrm{edge}$ is rather small, a noisy demonstration of the non-topological phase gate can be readily achieved even for $T_\textrm{edge}\sim\qty{1}{\kelvin}$.
Moreover, this temperature constraint is absent in the gate-controlled setup and in the discrete-time phase gate scheme described in the next section.

\subsection{Discrete-time scheme}\label{app:discrete_phase_gate}

We now elaborate upon the discrete-time phase gate scheme wherein interferometry takes place one anyon at a time.
The interferometry process, as depicted in Figs.~\ref{fig:phase_gate_3} and \ref{fig:phase_gate_electric}, acts upon the logical qubit as
\[
    \begin{aligned}
        \rho &\rightarrow \tilde{\rho} = \frac{\hat{O}\rho\hat{O}^\dagger}{\Tr\left[\hat{O}\rho\hat{O}^\dagger\right]}, \\
        \hat{O} &= \cos\left(\frac{\theta}{2}\right) I + \sin\left(\frac{\theta}{2}\right) e^{i\phi} Z.
    \end{aligned}
\]
Here, $\theta$ and $\phi$ parameterize the relative amplitude and dynamical phase difference between the two interfering paths, respectively.
In Sec.~\ref{sec:phase_gates}, we describe how the phase $\phi$ may be tuned in the tunnel-junction and gate-controlled setups.
The logical operator $Z$ appearing in $\hat{O}$ reflects the braid phase accumulated by probe anyon $\sigma_p$ encircling qubit anyons $\sigma_1$ and $\sigma_2$.
In particular, $Z$ is diagonal in the basis where $\sigma_1$ and $\sigma_2$ have definite fusion channel $I$ or $\psi$, for which the braid phase is $+1$ or $-1$, respectively.
For a generic initial state $\rho$, the resulting matrix elements are
\begin{equation}
    \begin{aligned}
        \tilde{\rho}_{00/11} &= \rho_{00/11} \frac{1 \pm \sin\theta\cos\phi}{1 + \sin\theta\cos\phi(\rho_{00} - \rho_{11})},\\
        \tilde{\rho}_{01} &= \rho_{01} \frac{\cos\theta + i\sin\theta\sin\phi}{1 + \sin\theta\cos\phi(\rho_{00} - \rho_{11})}.
    \end{aligned}
    \label{eq:discrete_phase_gate}
\end{equation}
As seen above, the operator $\hat{O}$ is non-unitary for general values of $\theta$ except for when the dynamical phases of the two paths differ by exactly $\phi = \pm\pi/2$.
Precisely at $\phi = \pm \pi/2$, this reduces to a simple phase gate $R_Z(\theta)$, yielding the final state
\[
    \tilde{\rho} = R_Z(\theta) \rho R_Z^\dagger(\theta) = \begin{pmatrix} \rho_{00} & \rho_{01}e^{i\theta} \\ \rho_{10} e^{-i\theta} & \rho_{11} \end{pmatrix}.
\]
Then to achieve a desired gate angle (e.g., $\theta=\pi/4$ for the magic gate), the probability of the probe anyon taking one path over the other must be tuned.

Moving away from the unitary limit introduces a finite degree of decoherence.
Consider a small phase error $\phi = \tfrac{\pi}{2} + \epsilon$ with $\abs{\epsilon} \ll 1$ such that $\hat{O} = \rotGate{Z}{\theta} - \epsilon \sin(\theta/2) Z + \mathcal{O}(\epsilon^2)$.
This yields a corresponding correction to the final state $\tilde \rho = \rotGate{Z}{\theta}\rho R_{Z}^\dagger(\theta) + \Delta \tilde \rho$, where
\[
    \Delta \tilde \rho = 2\sin\theta \left[-\rho_{00}\rho_{11} Z + (\rho_{00}-\rho_{11})(X\cos\theta - Y\sin\theta)\right].
\]
Notably, this correction includes both diagonal and off-diagonal terms, in contrast to the purely off-diagonal decoherence of the continuous-time scheme in Sec.~\ref{app:continuous_phase_gate}.

When the initial state lies in the equatorial plane of the Bloch sphere (i.e., $\rho_{00} = \rho_{11}$), the leading order effect of phase error $\epsilon \neq 0$ is an induced polarization of the qubit.
Namely, $\Delta\tilde\rho = -\tfrac12\sin(\theta) Z$ is entirely diagonal.
If we further require that $\rho$ is a pure state ($\abs{\rho_{01}} = 1/2$), then the action of $\hat{O}$ is described by an effective unitary,
\[
    U_\textrm{eff} = \rotGate{X}{\frac{\pi}{2}-\phi} \rotGate{Z}{\theta},
\]
such that $\tilde\rho = U_\textrm{eff} \rho U^\dagger_\textrm{eff}$.
Here the $R_X$ gate captures the polarization induced by the interferometry.
Such polarization errors can be mitigated by a two-step procedure akin to a spin-echo wherein one first performs the interferometry with parameters $(\theta, \phi)$ and then again with parameters $(\theta, \pi-\phi)$.
In the tunnel-junction setup, this phase reversal could be realized by inducing the dynamical phase $\phi$ on the other hole (by modifying the hole size) after pinching and applying a $Z$-gate by a braiding operation.
Similarly, we can perform this tuning of $\phi$ in the gate-controlled setup via the Aharonov-Bohm effect by altering the magnetic flux piercing the interferometry loop.
The total operator applied by this process is now unitary and for $\phi = \pi/2 + \epsilon$ is simply $R_Z(2\theta) + \mathcal{O}(\epsilon^2)$.
While this discrete-time phase gate scheme still requires careful tuning of parameters to achieve the desired phase gate, it yields distinct error channels which we have shown may be controlled and partially mitigated.
Moreover, by eliminating the spin liquid edge from the process, it alleviates the temperature constraints for the tunnel-junction setup discussed in Sec.~\ref{app:continuous_phase_gate}.

\end{appendix}
%%%%%%%%%%%%%%%%%%%%%%%%%%%%%%%%%%%%%%%%%%%%%%%%%%%%%%%%%%%%%%%%%%%%%%%%%%%%

\end{document}